\documentclass[11pt]{amsart}
\usepackage{amsaddr}
\usepackage{graphicx,amssymb,amsmath,amsthm}

\usepackage[utf8]{inputenc}
\usepackage{epstopdf}
\usepackage{bm}
\usepackage[numbers]{natbib}
\usepackage{color}
\usepackage{array}
\usepackage{url,hyperref}	

\usepackage{graphicx,amsmath,amsfonts}
\usepackage{epstopdf}

\usepackage{lipsum}
\usepackage{kantlipsum} 

\calclayout

\usepackage{amsfonts}
\usepackage{graphicx}
\usepackage{epstopdf}
\usepackage{algorithmic}
\usepackage{bm} 

\usepackage{amsmath}
\usepackage{amssymb}
\usepackage{amsfonts}
\usepackage{amsthm}
\usepackage{bbm}
\usepackage{bm}
\usepackage{graphicx}
\usepackage{setspace}
\usepackage{booktabs}

\usepackage{amsopn}

\newcommand{\LK}{\mathcal{L}}
\newcommand{\K}{\mathcal{K}}
\newcommand{\Kapprox}{\rm{K}}

\newcommand{\R}{\mathbb{R}}
\newcommand*{\Id}{\operatorname{Id}}
\newcommand{ \E}{\mathcal{E}_{\rm{DMD}}}
\newcommand{ \N}{\mathcal{N}_{\rm{DMD}}}
\def \O{\mathcal{O}}

\newcommand{\C}{\mathbb{C}}

\usepackage[colorinlistoftodos]{todonotes}
\newcounter{gagcomment}
\usepackage{amsfonts}

\title[Detecting regime transitions using dynamic mode decomposition]{Detecting regime transitions in time series using dynamic mode decomposition}
\author{Georg A. Gottwald and Federica Gugole}
\address{School of Mathematics and Statistics, University of Sydney, NSW 2006, Australia}
\address{Center for Earth System Research and Sustainability (CEN), Meteorological Institute, University of Hamburg, Germany}
\email[G. A. Gottwald and F. Gugole]{georg.gottwald@sydney.edu.au {\rmfamily and} federica.gugole@uni-hamburg.de}

\begin{document}

\begin{abstract}
We employ the framework of the Koopman operator and dynamic mode decomposition to devise a computationally cheap and easily implementable method to detect transient dynamics and regime changes in time series. We argue that typically transient dynamics experiences the full state space dimension with subsequent fast relaxation towards the attractor. In equilibrium, on the other hand, the dynamics evolves on a slower time scale on a lower dimensional attractor. The reconstruction error of a dynamic mode decomposition is used to monitor the inability of the time series to resolve the fast relaxation towards the attractor as well as the effective dimension of the dynamics. We illustrate our method by detecting transient dynamics in the Kuramoto-Sivashinsky equation. We further apply our method to atmospheric reanalysis data; our diagnostics detects the transition from a predominantly negative North Atlantic Oscillation (NAO) to a predominantly positive NAO around 1970, as well as the recently found regime change in the Southern Hemisphere atmospheric circulation around 1970.
\end{abstract}



\maketitle


\section{Introduction}
%
%
Being able to determine whether a dynamical system is evolving on an attractor exhibiting essentially equilibrium dynamics, or whether it exhibits transient dynamics is of utmost importance. When a dynamical system undergoes a regime transition, it may have a dramatic impact, with often major societal and economic consequences.  Is a system exhibiting rare but natural equilibrium fluctuations, or is the system transitioning, possibly to a new equilibrium state with possibly very different statistical behaviour? Such regime changes may be reflected in clearly separable dynamic states, akin to the transition of a particle in a double well potential under small noise, leading to a bimodal probability distribution. In the climate system, which by and large supports unimodal probability distributions for quantities such as pressure, however, the regime changes may appear more subtly \cite{Wirth01}, and scientists have devised methods to detect transitions, often tailored to particular dynamic variables measuring their statistics such as, for example, the frequency of atmospheric blocking events. We develop here a computationally cheap and easily implementable method which aims at answering this question when only a time series is given. We consider dissipative dynamical systems which evolve, when in equilibrium, on an attractor the dimension of which is smaller than the full state space dimension. We consider the case of an externally driven transition to a new attractor as well as transient non-equilibrium dynamics off the attractor. We use the term equilibrium here as being a statistically stationary state characterized by a stationary measure which is supported on the attractor.\\ 

For low dimensional deterministic systems experiencing bifurcations, critical slowing down near the perturbation and an associated increase in the variance can be used to detect transitions and provide early warning indicators for such transitions (see, for example, \cite{Kuehn11} and references therein). More recently, network theory and changes in the underlying network topology \cite{vanderMheenEtAl13,FengEtAl14,TiribassiEtAl14} as well as methods analyzing the point spectrum of the transfer operator \cite{ChekrounEtAl14,TantetEtAl15,TantetEtAl18a,TantetEtAl18b} were used to detect transitions in high-dimensional systems. We use here the framework of the Koopman operator, the formal adjoint of the transfer operator, and of dynamic mode decomposition (DMD) \cite{Mezic13,KutzEtAl,KlusEtAl18}. The Koopman operator is an abstract concept in dynamical systems theory which encodes the dynamics of a dynamical system, propagating observables from one instance of time to another instance of time \cite{LasotaYorke}. A computationally cost-effective algorithm attempting to compute a finite-dimensional approximation of the Koopman operator is given by DMD, which was proposed by Schmid {\em{et al}} \cite{Schmid10,SchmidEtAl11}. The connection between DMD and the Koopman operator has been made first by Rowley {\em{et al}} \cite{RowleyEtAl09}, and has lead to further fruitful extensions \cite{TuEtAl14,WilliamsEtAl15,NoackEtAl16}. DMD distills dynamically relevant structures, the so called Koopman modes together with their temporal oscillation periods and/or their growth rates. Compared to other model reduction techniques such as principal orthogonal decomposition (POD), which optimally reconstruct data by maximizing the energy contained in the POD modes, DMD decomposes the dynamics according to its local in time oscillatory behaviour.  Connections with POD and DMD are discussed in \cite{Schmid10,TuEtAl14}. There are also connections of DMD with other model reduction methods such as the eigensystem realization algorithm \cite{JuangPappa85} and linear inverse modelling \cite{Penland89,PenlandMagorian93}. A detailed overview about DMD and its connections with other dimension reduction techniques can be found in \cite{TuEtAl14}.  DMD has been used in many areas of science to unravel dynamical features such as instabilities and bifurcations (see, for example, \cite{BudisicEtAl12,Bagheri13,KutzEtAl,KlusEtAl16,KlusEtAl18}). In particular, DMD allows for an approximate  reconstruction of the dynamics, locally in time, by the eigenmodes of the Koopman operator. It is important to realize though that DMD in general does not allow for a faithful approximation of the Koopman operator and its spectral properties, and that the reconstruction of the dynamics is only possible, if at all, on finite time intervals \cite{TuEtAl14,WilliamsEtAl15}. For chaotic dynamical systems evolving on a lower-dimensional attractor, it suffices to express the dynamics with a finite number of Koopman modes; the number of modes needed being proportional to the attractor dimension. The main idea of this work is the following: Typically in dissipative systems non-equilibrium states off the attractor experience fast relaxation towards the attractor on time scales much faster than those associated with the equilibrium dynamics evolving on the attractor. Resolving this fast relaxation requires a sufficiently fine temporal resolution, which may not be given in the time series under consideration. Hence, contrary to the case of equilibrium dynamics on the attractor, during transient dynamics the reconstruction may be poor for sampling times which were appropriate to resolve the equilibrium dynamics. Moreover, in equilibrium the dynamics evolves on a lower dimensional attractor, whereas transient dynamics feels the full state space dimension. This may cause poor the reconstruction of transient dynamics by a finite number of Koopman modes, which would be sufficient to capture the equilibrium dynamics. We propose here a simple and computationally cheap method to detect non-equilibrium dynamics, such as transients or qualitative regime changes of the attractor, by monitoring the reconstruction error of DMDs. We illustrate the effectiveness of our method using the Kuramoto-Sivashinsky equation for which the existence of a finite-dimensional attractor has been proven \cite{Temam}. Furthermore, we apply DMD to atmospheric reanalysis data \cite{NCEPdata}. Our method is able to detect the transition from a predominantly negative North Atlantic Oscillation (NAO) phase to a predominantely positive phase in the early 1970s, which was shown to be related to a change in forecast skill  \cite{HurrellVanLoon97,WoollingsEtAl15,WeisheimerEtAl17}. The reconstruction error also detects the regime change, attributed to an increased  ${\rm{CO}}_2$ concentration, of the Southern Hemisphere atmospheric dynamics around 1970 from a regime with more intense baroclinicity to a regime of more zonal barotropic dynamics \cite{FrederiksenFrederiksen07,FranzkeEtAl15,FreitasEtAl15}.\\

The paper is organized as follows. In Section~\ref{sec.DMD} we briefly present the method of dynamic mode decomposition and propose our diagnostic for the detection of transients using the reconstruction error. In Section~\ref{sec.KS} we apply our method to detect transitions and regime changes in the Kuramoto-Sivashinsky equation. In Section~\ref{sec.NCEP_NAO} we apply our method to analyze the regime change of the Northern Hemisphere atmospheric circulation dynamics and the NAO in reanalysis data. The regime change of the Southern Hemisphere atmospheric circulation dynamics is analyzed using reanalysis data in Section~\ref{sec.NCEP}. We conclude with a discussion in Section~\ref{sec.discussion}.


\section{Dynamic mode decomposition and Koopman modes}
\label{sec.DMD}
In the following we introduce the Koopman operator and the computationally cheap dynamical mode decomposition algorithm for the approximation of the Koopman operator and its eigenfunctions.

Let us consider a dynamical system
\begin{align}
\dot x = f(x)
\label{e.ds}
\end{align}
with $x(0)=x_0$ for $x\in \R^d$. Provided there is a unique solution of this initial value problem, we can introduce the flow map $\varphi_t$ and write $x(t)=\varphi_t(x_0)$. Consider observables $\psi(x)$. Observables are propagated in time according to $\psi(\varphi_t(x))$ which we express as
\begin{align}
\K_t \psi = \psi(\varphi_t(x)),
\end{align}
where the propagator $\K$ is termed the Koopman operator. Applying the chain rule one can verify that continuously differentiable  observables satisfy the following linear partial differential equation
\begin{align}
\partial_t v(x,t) &= \LK v
\label{e.Lt}
\\
\nonumber
v(x,0)&=\psi(x),
\end{align} 
with the generator $\LK=f(x)\cdot\nabla$. We formally solve this Cauchy problem and write $v(x,t)=\left(e^{\LK t}\psi\right)(x) = \psi(\varphi_t(x))$, and identify 
\begin{align*}
\K_t=e^{\LK t}.
\end{align*}
The notion of the Koopman operator can be extended to bounded (not necessarily continuously differentiable) observables for which the limit defining the generator
\begin{align}
\LK \psi =\lim_{t \to 0} \frac{e^{\LK t}\psi -\psi}{t}
\label{e.LieAlgebra}
\end{align}
exists. We remark that this can be readily extended for stochastic differential equations. For details the reader is referred to \cite{LasotaYorke,PavliotisStuart}. In order to express numerically these infinite-dimensional operators, one typically considers these operators as infinitely large matrices and constructs finite-rank representations. This, however, is strictly admissible only for compact operators, and it has been proven only for certain classes of dynamical systems; we refer the interested reader to \cite{Bollt} and references therein.

Let us now describe DMD and how the Koopman operator and its eigenfunctions can be approximated given a set of observations. We follow here the exposition provided in \cite{TuEtAl14,KutzEtAl}. We are given snapshots $X$
\begin{align*}
X = 
\begin{pmatrix}
   \vert & \vert &  & \vert\\
   x_1 & x_2 & \cdots & x_m \\
   \vert & \vert &  & \vert
\end{pmatrix} 
\end{align*}
with $x_k=x(t_k)\in\R^n$. Successive vectors $x_k$ have evolved from $x_{k-1}$ under the dynamics for some, not necessarily small time interval $\Delta t$, i.e. $x_k = \K_{\Delta t} x_{k-1}$. We are further given snapshots $X^\prime$
\begin{align*}
X^\prime = 
\begin{pmatrix}
   \vert & \vert &  & \vert\\
   x_1^\prime & x_2^\prime & \cdots & x_m^\prime \\
   \vert & \vert &  & \vert
\end{pmatrix} 
\end{align*}
with $x_k^\prime=x(t_k+\delta t)\in\R^n$. The variables $x$ may denote the state variables, or any observable of them\footnote{DMD was originally introduced in the case when $x$ denotes the state variables.}. Successive vectors $x_k^\prime$ have evolved from $x_{k}$ under the dynamics for a time $\delta t$, i.e. $x_k^\prime = \K_{\delta t} x_{k}$. The size of $\delta t$ determines the accuracy of the reconstructed dynamics and should be chosen sufficiently small, as can be seen from (\ref{e.LieAlgebra}).
In DMD the Koopman operator $\K_{\delta t}$ is approximated by a least square fit from $X^\prime = \K_{\delta t} X$ as
\begin{align}
\Kapprox = X^\prime X^\dagger,
\label{e.XXpseudo}
\end{align}
where $X^\dagger$ denotes the pseudo-inverse of $X$.  The finite dimensional approximation of the Koopman operator (\ref{e.XXpseudo}) may suggest that it is tacitly assumed that the dynamics is linear. This is not the case, only the infinite dimensional dynamics (\ref{e.Lt}) determining the Koopman operator (or rather its generator) is linear. Recall that the Koopman operator acts on functions $\psi(x)$; in the case when the observables are the actual state variables, we have $\psi(x)={\rm{id}}(x)$. If the observables span a sufficiently large space the linear least square fit (\ref{e.XXpseudo}) is indeed able to capture the full nonlinear dynamics (\ref{e.ds}) \cite{TuEtAl14}. 
We express $\Kapprox$ as
\begin{align*}
\Kapprox = X^\prime V\Sigma^{-1}U^\star,
\end{align*}
where the star denotes the complex conjugate transpose and we use a low rank $r\le m$ singular value decomposition of $X=U\Sigma V^\star$  with the proper orthogonal decomposition (POD) modes $U\in\C^{n\times r}$, $\Sigma\in\C^{r\times r}$ and $V\in \C^{m\times r}$ and $U^\star U=\Id$ and $V^\star V=\Id$. For computational efficiency $\Kapprox$ is projected onto the POD modes and we consider
\begin{align*}
\tilde \Kapprox = U^\star \Kapprox U = U^\star X^\prime V \Sigma^{-1}.
\end{align*}
Performing an eigendecomposition of $\tilde\Kapprox$ with $\tilde\Kapprox W = W\Lambda$, eigenmodes of the approximation of the Koopman operator $\K_{\delta t}$ are expressed as
\begin{align*}
\Phi = X^\prime V \Sigma^{-1}W\Lambda^{-1},
\end{align*}
 satisfying $\Kapprox \Phi=\Phi\Lambda$. Note that eigenmodes of $\tilde \Kapprox$ are given by $\tilde\Phi = UW$. Whereas the eigenmodes $\tilde \Phi$ are orthonormal, this is not the case for $\Phi$.  Introducing $\omega_j = \ln \lambda_j/\delta t$, where we use the principal branch of the logarithm, the snapshots $x_k=x(t_k)$ are now approximated by the DMD-reconstructed field as
\begin{align}
x(t_{k+1}) \approx \sum_{j=1}^r \phi_je^{\omega_j k\Delta t}\, b_j + {\rm{c.c.}} =\Phi\exp(\Omega k\Delta t) \, b + {\rm{c.c.}},
\label{e.DMDrecon}
\end{align} 
where $b=\Phi^\dagger x(t_1)$ denotes the initial coefficients, $\Omega \in \mathbb{C}^r$ has components $\omega_j$, and ${\rm{c.c.}}$ denotes the complex conjugate. 

Tu {\em{et al}} \cite{TuEtAl14} showed that DMD only provides a good approximation of the Koopman operator and its spectral properties provided the data are sufficiently diverse -- i.e. the sampling time $\Delta t$ is sufficiently large to ensure sufficient diversity and a large range of $X$ -- and provided the observables are sufficiently rich, in the sense that their span contains the eigenfunctions of the Koopman operator. It is pertinent to mention that it is {\em{a priori}} not possible to determine if given observables are sufficient to span the eigenfunctions of the Koopman operator, and it is typically not to be expected that a given set of observables suffices. To address this issue the extended dynamic mode decomposition (EDMD) \cite{WilliamsEtAl15,KordaMezic18} and DMD algorithms based on Hankel matrices of data \cite{ArbabiMezic17,BruntonEtAl17} were proposed and were shown to provide a better approximation of the spectral properties of the Koopman operator than DMD.\\ 
 
We focus here only on {\em{reconstruction}} of the observations by the Koopman modes and not on forecasting. Hence the question whether the span of the observables is sufficiently large to contain all Koopman eigenmodes is less relevant here. A DMD reconstruction involves the reconstruction of the whole spatio-temporal evolution for a finite time window of $m\Delta t$ time units with $m$ snapshots (where the number of snapshots $m$ is typically much smaller than the total number of observations which are available), and we do not employ the DMD approximation (\ref{e.DMDrecon}) past the observations given in the time window $m\Delta t$. We define the reconstruction error 
\begin{align}
\label{e.E}
\E(t_{k},r) = \frac{1}{m} \sum_{l=0}^{m-1} 
\| x(t_{k+l}) - \Phi\exp(\Omega l \Delta t) b_k - {\rm{c.c}}\|,
\end{align} 
with $b_k = \Phi^\dagger x_{k}$. We are given data sets $X$ and $X^\prime$ with $M\gg m$ snapshots collected at intervals of length $\Delta t$. We split these time series into windows of temporal length $m\Delta t$ and reconstruct the field by means of DMD for each of the windows. In the following Section we consider $\delta t=\Delta t$ for our analyses, which is the typical situation in time series analysis. However in case of artificially produced data it may be advantageous to choose $\delta t \ll \Delta t$ to achieve better DMD analyses. Since the aim of this work is the detection of a breakdown of the DMD reconstruction and we are not interested in the optimal way to perform DMD analysis, we use $\delta t =\Delta t$. Typically, the matrices $X$ and $X^\prime$ are skinny and tall with the number of snapshots $m$ smaller than the dimension $n$ of a snapshot $x$. The reconstruction error is local in time and we reconstruct $M/m$ time windows of temporal length $m\Delta t$. If the dynamics is in equilibrium at time $t_k$, evolving on an attractor with dimension $d_{\rm{a}}\ll d$ we expect that (for a sufficiently small sampling time) the reconstruction error $\E(t_k,r)$ is large for $r\le r_a$ with $r_a\sim\O(d_{\rm{a}})$, and then rapidly decreases for larger values of $r$. If the dynamics is, on the other hand, not in equilibrium at time $t_k$ and experiences the full state space dimension $d$, the value of $r$ needed for good reconstruction may be larger than the value $r$, which was chosen based on previous knowledge of the equilibrium dynamics. More importantly, the fast attraction towards the attractor with rate $\lambda_{\rm{relax}}$ may not be resolved by $\delta t$ if $\delta t \gg 1/\lambda_{\rm{relax}}$, implying a bad reconstruction with large values of $\E(t_k,r)$. We will see that the latter point is crucial in identifying transitions.


\section{Applications}
\label{sec.appl}
We now explore how the DMD reconstruction error can be used to detect transitions and regime changes. We start with artificially generated data from numerical simulations of a partial differential equation, before applying our method to confirm recent findings in regime changes and transitions in the atmospheric circulation of the Northern and the Southern Hemisphere.


\subsection{Detecting regime changes in the Kuramoto-Sivashinsky equation}
\label{sec.KS}
We first consider artificially generated time series obtained from a numerical simulation of the Kuramoto-Sivashinsky equation
\begin{align}
u_t + u u_x + \alpha u_{xx} + u_{xxxx} = 0.
\label{e.KS}
\end{align}
For fixed system length $L$ the Kuramoto-Sivashinsky equation becomes chaotic upon increasing the driving $\alpha$. The most unstable wave number is given through linearization around $u=0$ as $k_{\rm{max}} = \sqrt{\alpha/2}$ suggesting that the observed spatio-temporal patterns have around $(L/2\pi)\sqrt{\alpha/2}$ peaks.\\ We choose a fixed domain length $L=53.35$, and integrate the equation using a pseudo-spectral Crank-Nicolson scheme where the nonlinearity is treated with a second-order Adams-Bashforth scheme. We employ a discretization step of $dt=0.01$ and use $128$ spatial grid points. 

\begin{figure}[htbp]
	\centering
	\includegraphics[width = 0.7\columnwidth]{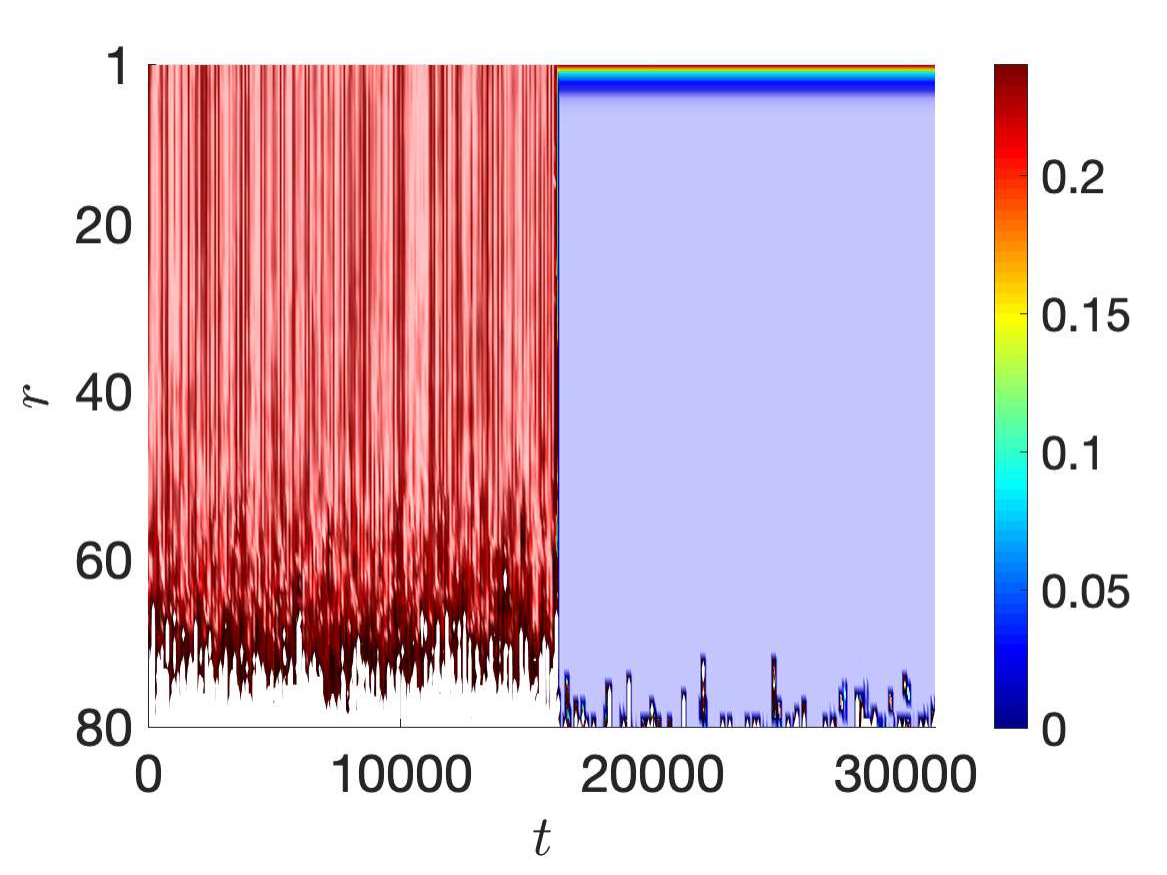}
	\caption{Reconstruction error $\E(t,r)$ for the Kuramoto-Sivashinsky equation (\ref{e.KS}) with $\alpha=2.53$ during transient dynamics from an initial condition off the attractor. The colouring was capped at $\E(t,r)=0.24$.}
	\label{f.KS_transient}
\end{figure}

\begin{figure}[htbp]
	\centering
	\includegraphics[width = 0.6\columnwidth]{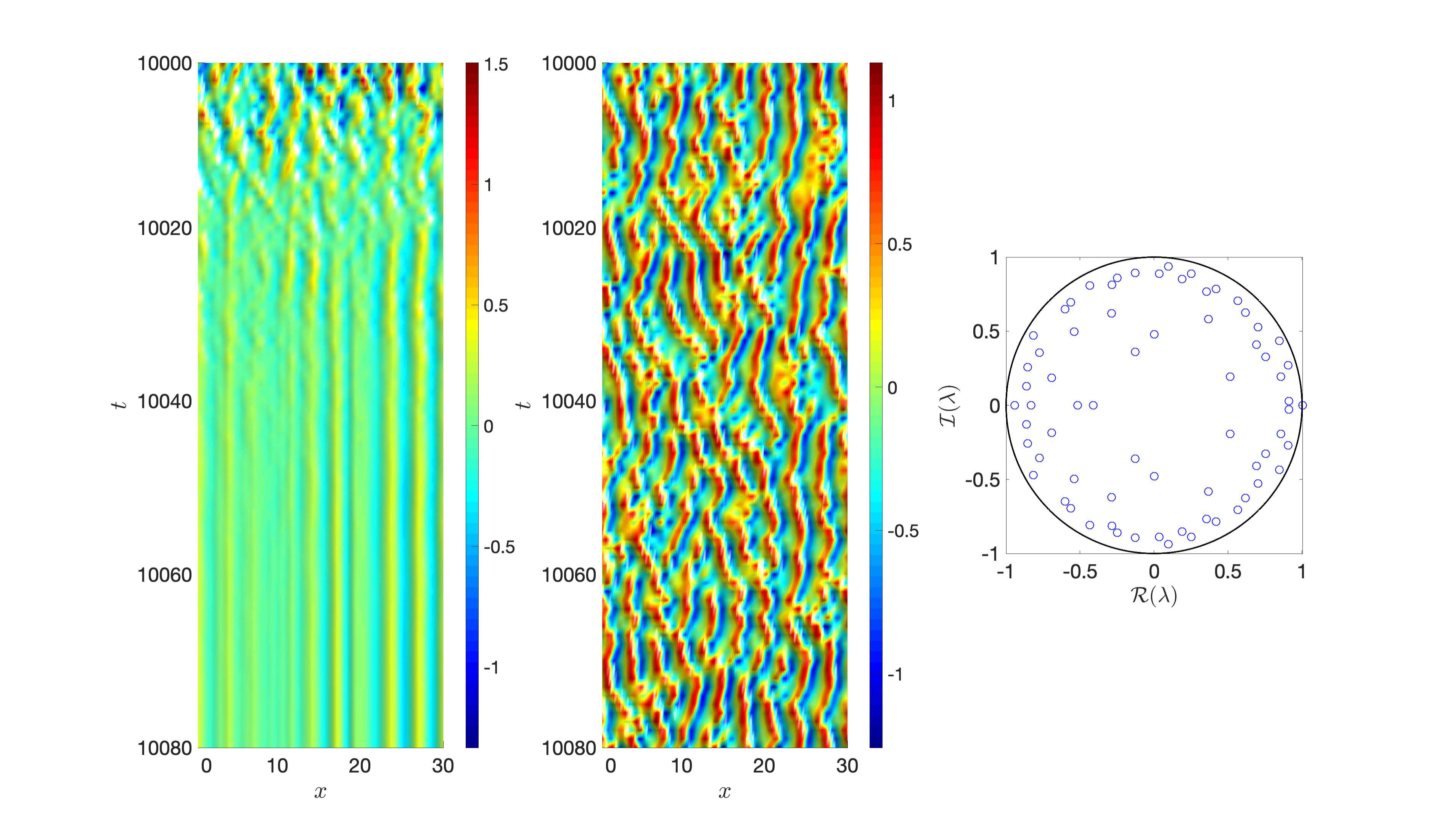}\\
	\vspace{0.1cm}
	\includegraphics[width = 0.6\columnwidth]{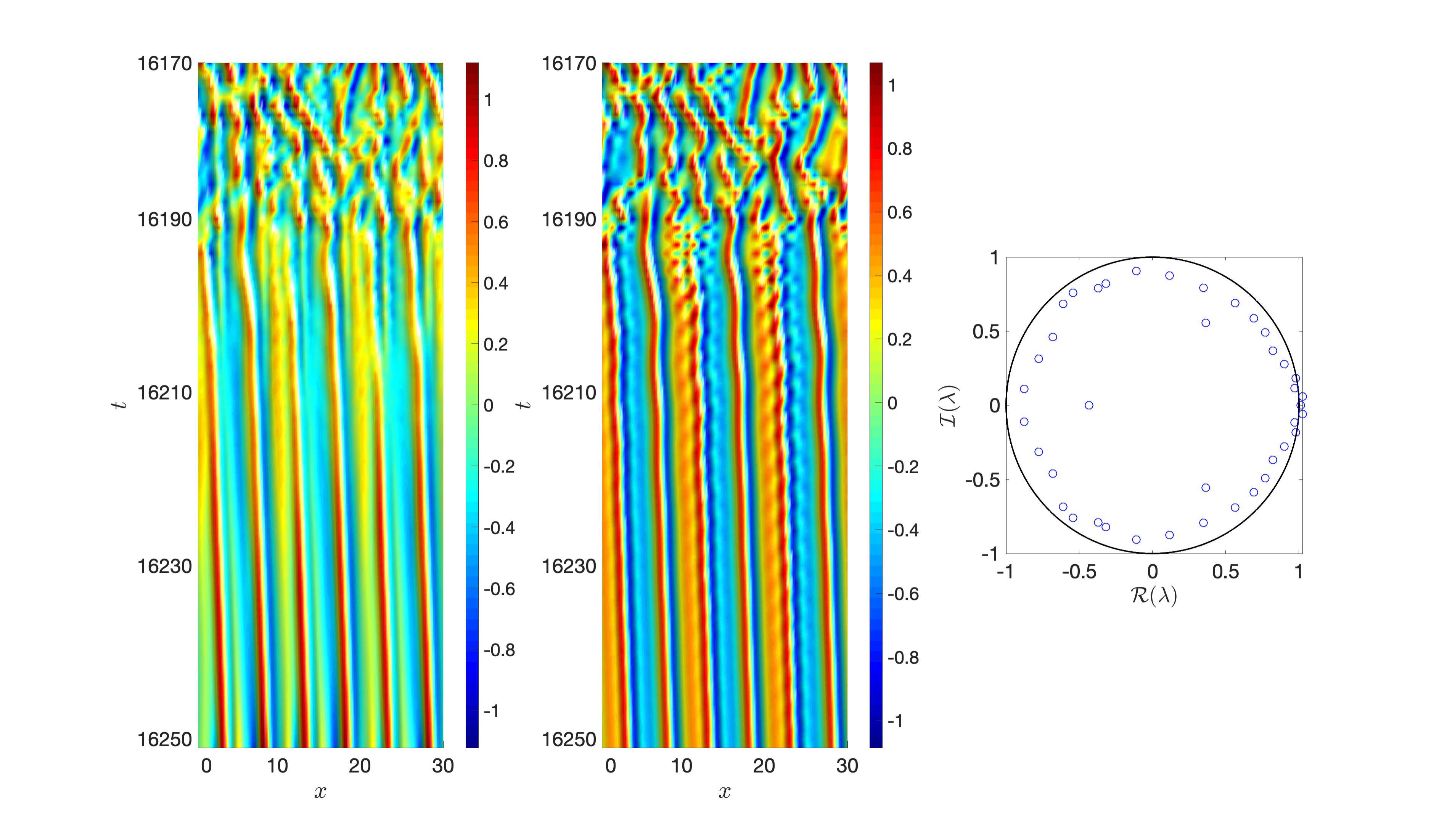}\\
	\vspace{0.1cm}
	\includegraphics[width = 0.6\columnwidth]{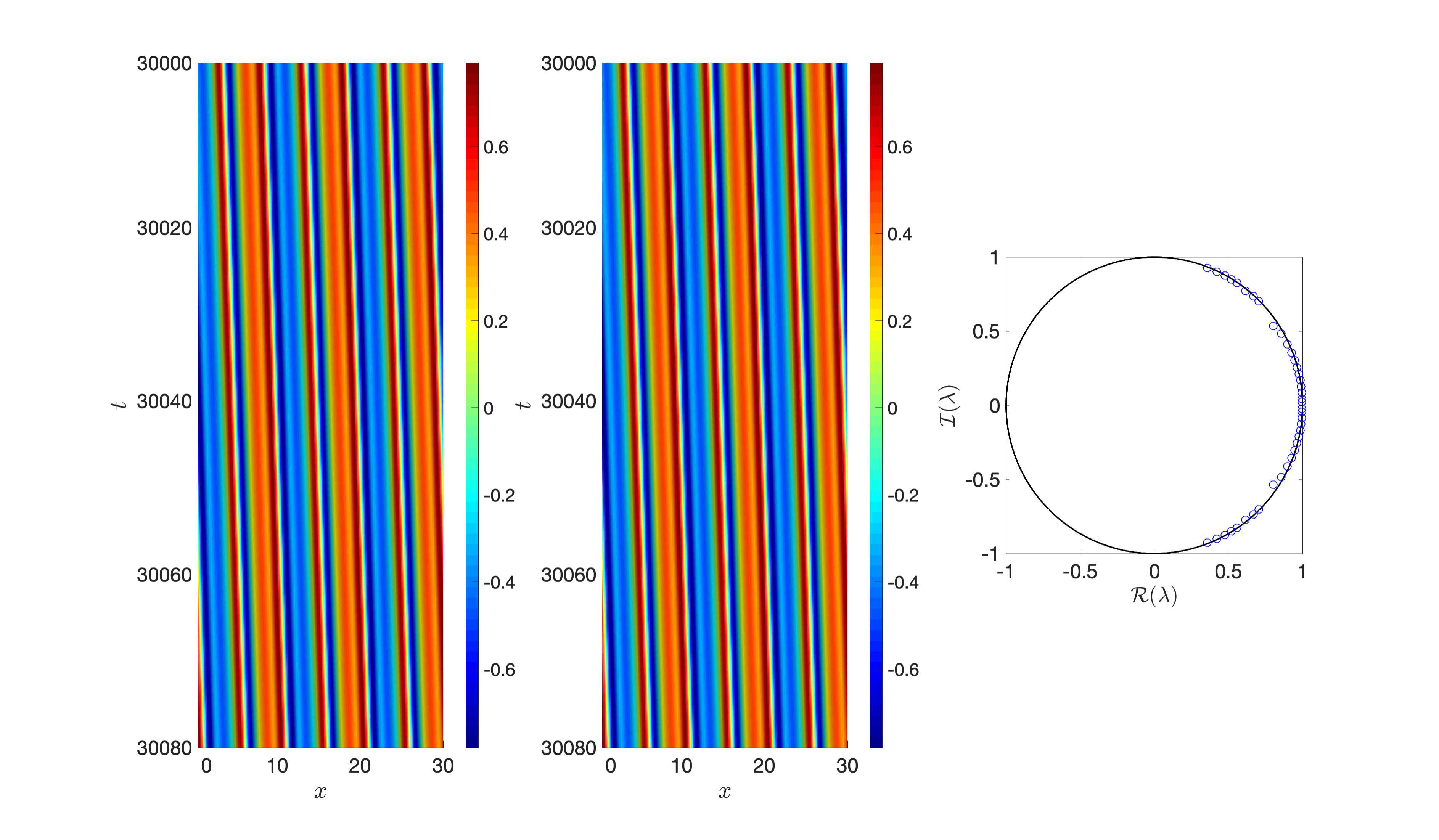}
	\caption{Waterfall plot of $u$ for the Kuramoto-Sivashinsky equation (\ref{e.KS}) with $\alpha=2.53$ (middle column) and for the associated DMD reconstruction with $m=80$ (left column). In the right column we show the DMD eigenvalues associated with the reconstruction. Top: Chaotic transitory dynamics at $t=10000$. For the DMD reconstruction we used $r=69$ corresponding to the minimal reconstruction error with $\E=0.224$. Middle: Just before relaxing on the limit cycle at $t=16170$. For the DMD reconstruction we used $r=40$ corresponding to a reconstruction error of $\E=0.614$.  Bottom: Regular limit cycle dynamics at $t=30000$. For the DMD reconstruction we used $r=7$ corresponding to a reconstruction error with $\E=4.8\times  10^{-5}$. }
	\label{f.KS_transient_Comparison}
\end{figure}

\begin{figure}[htbp]
	\centering
	\includegraphics[width = 0.5\columnwidth]{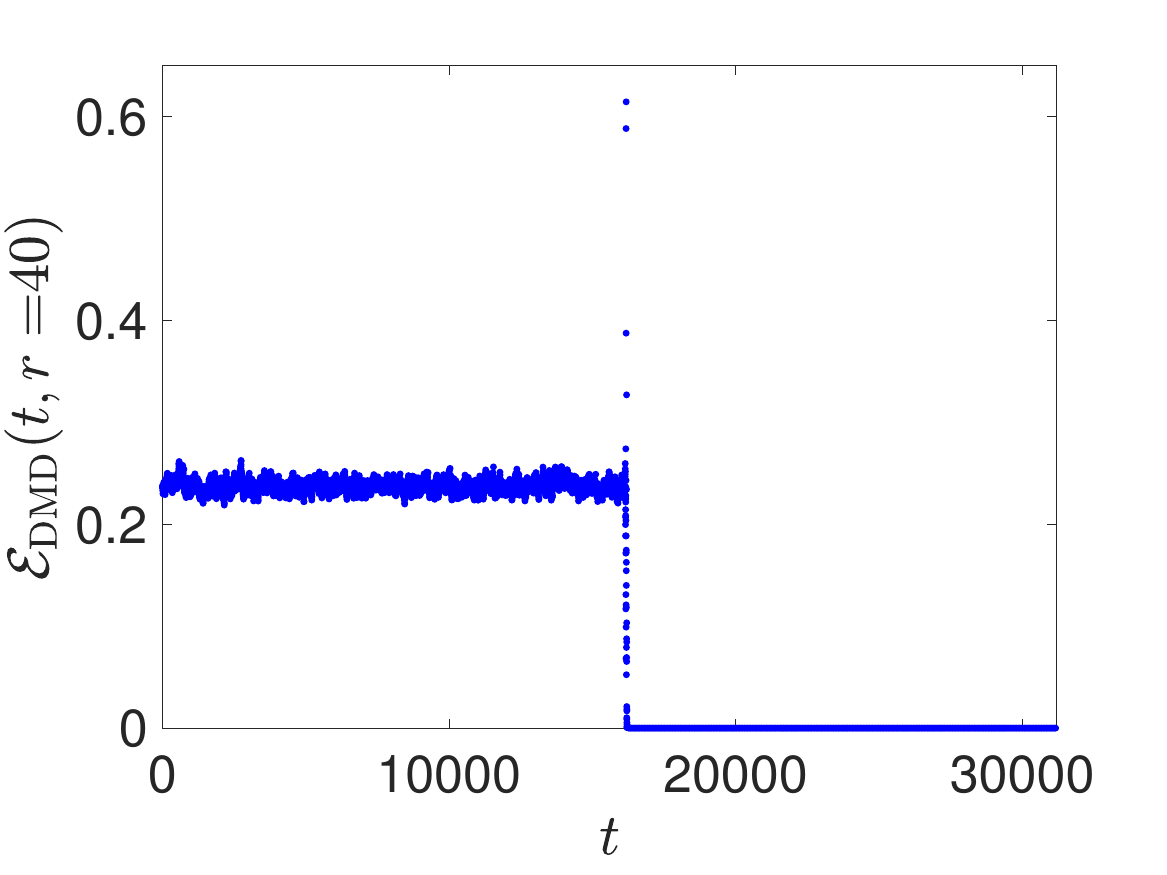}\\
	\vspace{0.1cm}
	\includegraphics[width = 0.5\columnwidth]{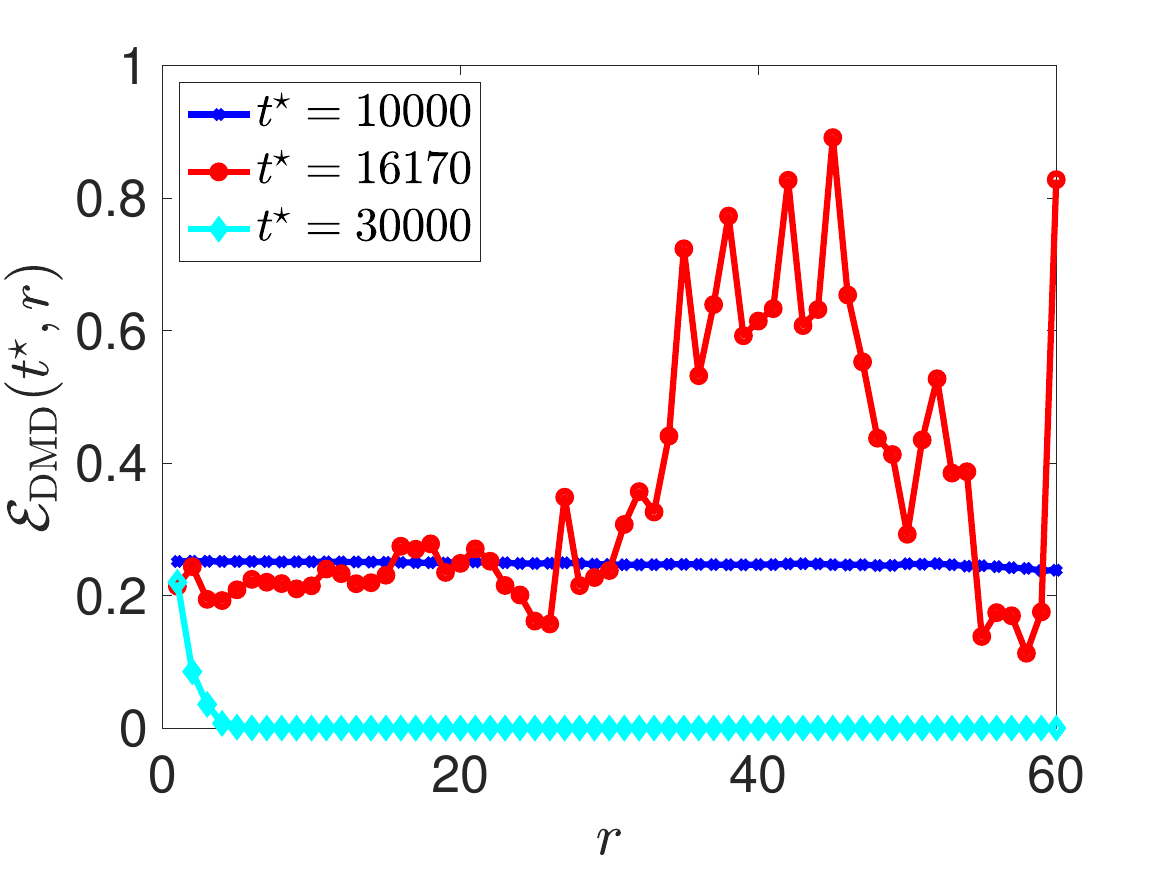}
	\caption{Reconstruction error $\E$ for the Kuramoto-Sivashinsky equation (\ref{e.KS}) with $\alpha=2.53$ during transient dynamics from an initial condition off the attractor. Top: $\E(t,r)$ at fixed $r=40$. Bottom: $\E(t^\star,r)$ for different fixed times $t^\star$ corresponding to chaotic transient dynamics ($t^\star=10000$), just before the relaxation to the limit cycle ($t^\star=16170$) and on the limit cycle ($t^\star=30000$).}
	\label{f.KS_transient_EDMDT_t}
\end{figure}

Our first experiment involves transient dynamics from an initial condition of $u(x,0)=u_0 \exp(-w(x-\tfrac{L}{2})^2)$ for $31000$ time units with $u_0=0.67$ and $w=0.62$ for $\alpha=2.53$ for which the dynamics is non-chaotic. The dynamics settles on a limit cycle with regular behaviour with period of $295.4$ time units around $16000$ time units. We choose $m=80$ snapshots to reconstruct the dynamics separated by $\delta t= \Delta t=1.0$. The DMD analysis is performed for the $128$-dimensional observable of the discretized field $u$ at every grid point. 
In Figure~\ref{f.KS_transient} we show the reconstruction error $\E(t,r)$. It is clearly seen that during the transient dynamics the reconstruction error is large but drops off significantly when the dynamics has settled down on the limit cycle near $t=16000$. Note that the number of POD modes required for an accurate reconstruction for the limit cycle dynamics is roughly $5$ and is smaller than the number of linearly unstable modes $\sqrt{\alpha} L/2\pi = 13$. Figure~\ref{f.KS_transient_Comparison} shows examples of DMD reconstructions and eigenvalues $\lambda$ for time  intervals of length $m \Delta t=80$ time units alongside their associated true fields for times $t=10000$, $t=16170$ and $t=30000$ time units, corresponding to the chaotic transient dynamics, the dynamics just before relaxing towards the limit cycle, and the regular limit cycle, respectively. Whereas the DMD reconstruction is almost indistinguishable from the true field on the limit cycle, it fails to reproduce the dynamics in the transitory region, where we employed $r=69$ for which the error was minimal. In the transitory region all eigenvalues $\lambda_i$ lie within the unit circle leading to an exponential decay in time. The mode which survives the longest, i.e. which is associated with the eigenvalue with largest real part, can be seen to have $9$ peaks, which corresponds to the wave number $\sqrt{\alpha/2} L/2\pi = 9.55$ of the most unstable linear wave. Just before relaxing to the limit cycle, the error is maximal. Here the eigenvalues are closer to the unit circle but the DMD analysis fails to reproduce the true dynamics. In Figure~\ref{f.KS_transient_EDMDT_t} we show how the reconstruction error for fixed $r=40$ evolves in time. We can identify three regions in terms of their reconstruction error: during transient chaotic dynamics the reconstruction error has values varying closely around $\E\approx 0.23$, before relaxing to the limit cycle where the reconstruction error is of the order of $10^{-20}$. These two regions are separated in time by a brief period around $t=16170$, during which the dynamics rapidly settles onto the limit cycle and during which reconstruction errors can exceed $\E= 0.6$. We further show in Figure~\ref{f.KS_transient_EDMDT_t} how the reconstruction error behaves as a function of $r$ for the three times $t=10000$, $t=16170$ and $t=30000$. Whereas the reconstruction error decays monotonically towards zero for the dynamics on the limit cycle at $t=30000$, the error does not significantly improve for the transient dynamics at $t=10000$ and at $t=16170$ upon increasing the number of Koopman modes, indicating that the dynamics can not be resolved using DMD for the given time series. For the abrupt transitory dynamics relaxing towards the limit cycle at $t=16170$, the error increases dramatically upon increasing $r$ beyond $r=30$. For the transient chaotic dynamics at $t=10000$ the error decreases until $r\approx 70$ (albeit not to acceptable values reflecting reliable reconstruction) before strongly increasing (not shown). For $r=m=80$, the DMD reconstruction yields reconstruction errors of the order of $10^{10}$ or higher for the whole duration of the transient dynamics. Such large reconstruction errors are also observed for isolated times during the regular dynamics on the limit cycle. These large reconstruction errors are visible as white regions in Figure~\ref{f.KS_transient}, and are caused by eigenvalues $\lambda_i$ outside the unit circle. For $m\le 60$ these badly estimated eigenvalues do not occur.

We argue that the reason for this bad reconstruction is the fast relaxation towards the attractor from the initial condition which was chosen far from the attractor. With the given sampling time of $\delta t=\Delta t=1$ the DMD analysis is able to recover the dynamics on the time window $m\Delta t=80$ time units from the observations, evidenced by the very low values of $\E$ for $t>16200$. For transitory dynamics, however, this sampling time is not sufficient to allow DMD to capture the dynamics, leading to a wrong estimation of eigenvalues and Koopman modes. To corroborate this argument, we have checked that for finely sampled observations with $\delta t=\Delta t=0.05$ one obtains very good reconstruction, also for the transient dynamics, for time windows of $5$ time units corresponding to $m=100$. We remark that, unlike for the sampling time $\delta t$, the reconstruction error is insensitive to the length of the time window for a wide range of $m$. We have tested that the baseline values of the reconstruction error, i.e. $\E\approx 0.23$ and $\E\approx 0$ for the transient and the regular limit cycle dynamics, respectively, as depicted in Figure~\ref{f.KS_transient_EDMDT_t}, do not change upon varying $m\in(10,100)$. These results suggest the following test for the detection of regime transitions: Given a current regime one can determine a sampling time $\delta t$ such that DMD is able to reconstruct the dynamics. Monitoring the reconstruction error for subsequent times, a transition is detected, if it involves faster time scales than those which can be resolved by the fixed sampling time $\delta t$.\\

\begin{figure}[htbp]
	\centering
	\includegraphics[width = 0.7\columnwidth]{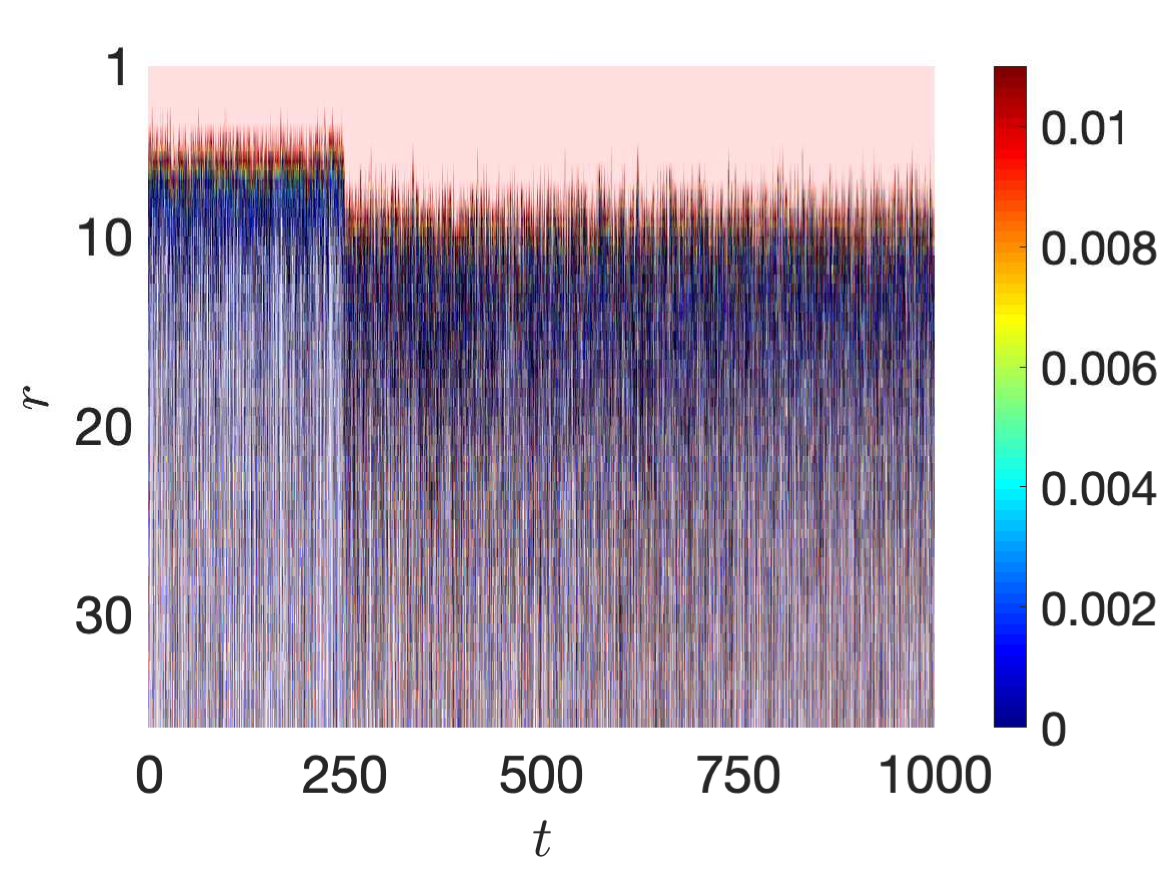}
	\caption{Reconstruction error $\E(t,r)$ for the Kuramoto-Sivashinsky equation (\ref{e.KS}) subject to an external change in $\alpha$ from $\alpha=3.16$ for $t\le 250$ to $\alpha=3.79$ for $t> 250$.}
	\label{f.KS_alpha}
\end{figure}
\begin{figure}[htbp]
	\centering
	\includegraphics[width = 0.5\columnwidth]{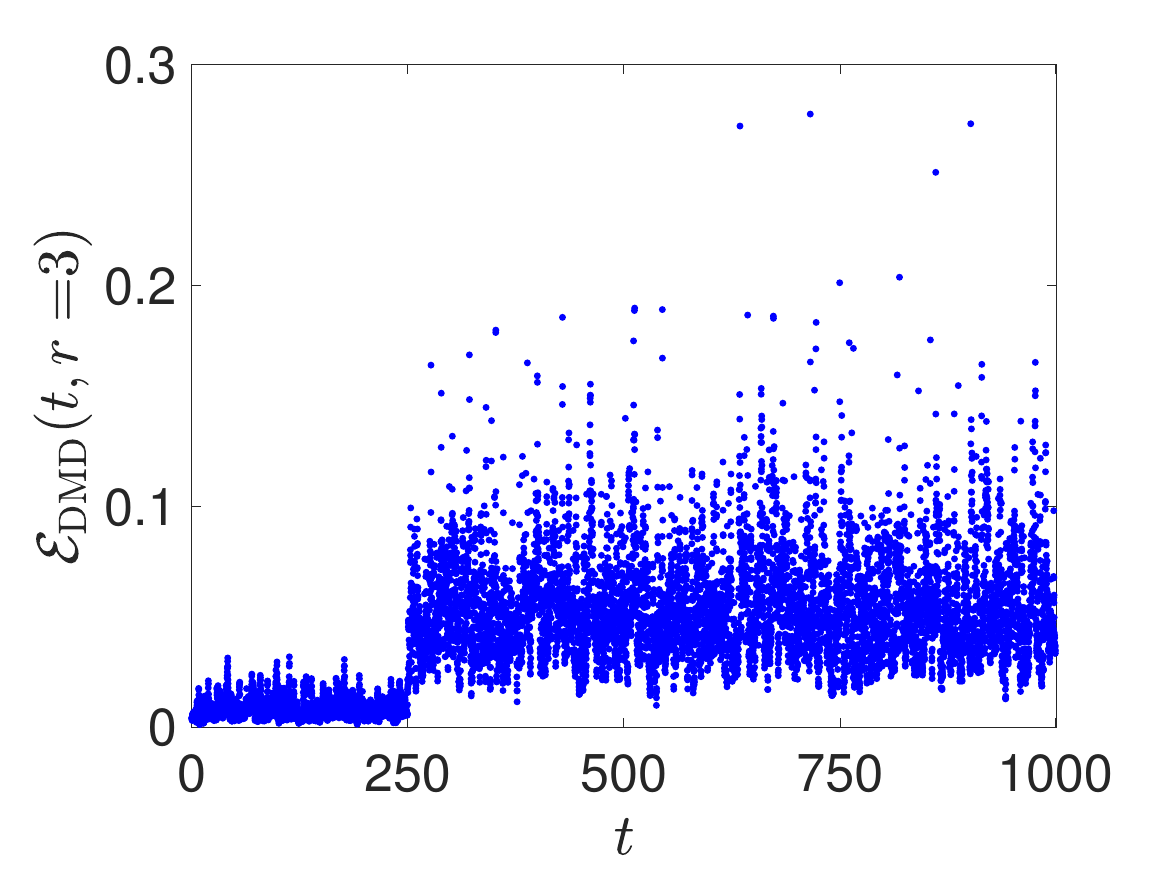}\\
	\vspace{0.1cm}
	\includegraphics[width = 0.25\columnwidth]{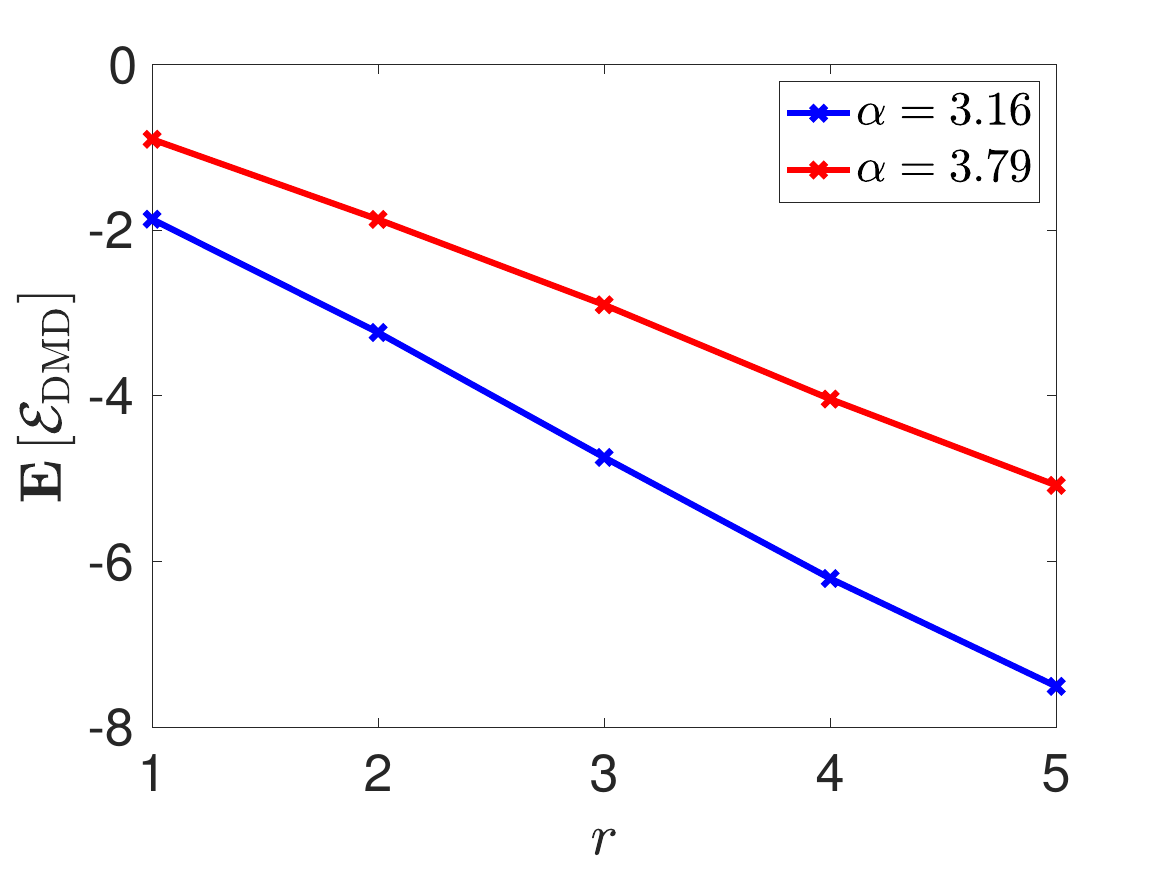}
	\includegraphics[width = 0.25\columnwidth]{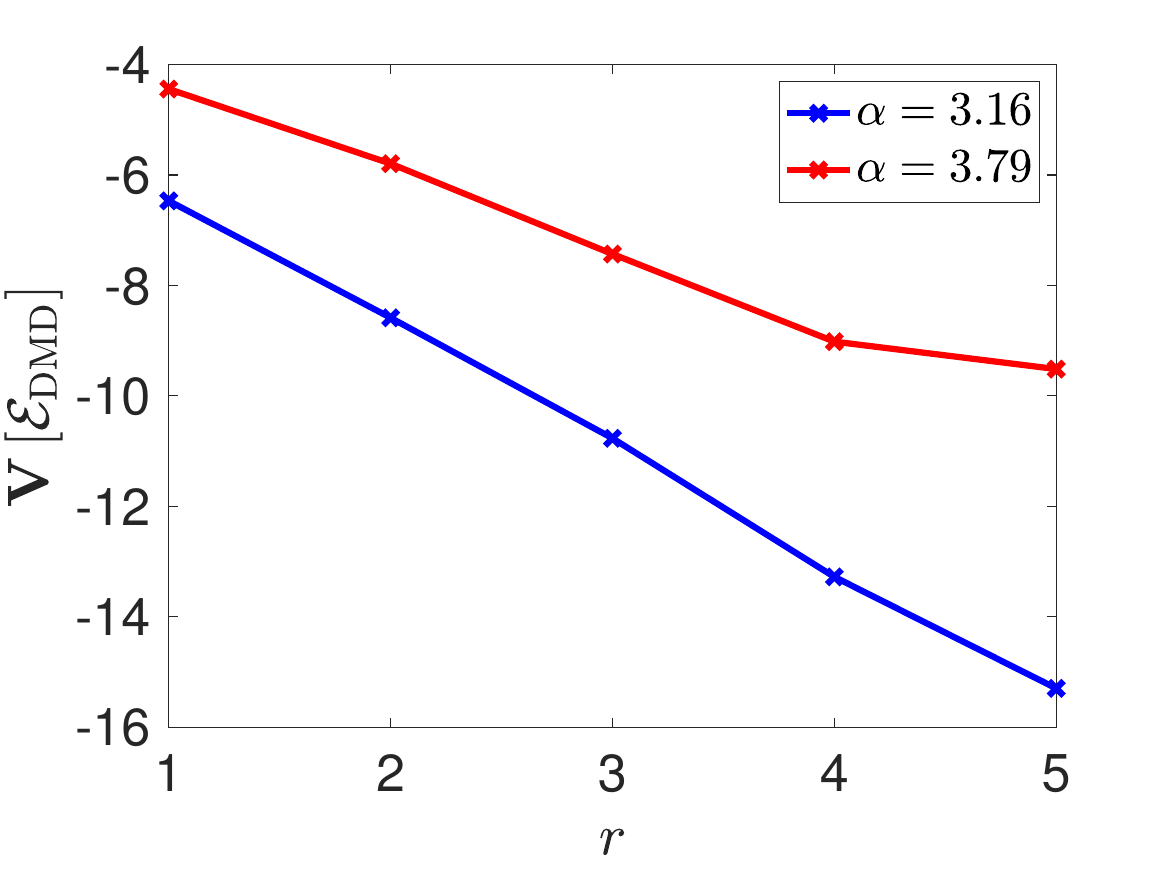}
	\caption{Top: Reconstruction error $\E(t,r)$ at fixed $r=3$ for the Kuramoto-Sivashinsky equation (\ref{e.KS}) subject to an external change in $\alpha$ from $\alpha=3.16$ for $t\le 250$ to $\alpha=3.79$ for $t>250$. Bottom: Mean (left) and variance (right) of the reconstruction error $\E(t,r)$ for the times  $t\le 250$ and for $t>250$, corresponding to $\alpha=3.16$ and $\alpha=3.79$, respectively.}
	\label{f.KS_alpha_meanvar}
\end{figure}

We further study the reconstruction error $\E(t,r)$ for dynamics exhibiting a regime change from one chaotic attractor to another chaotic attractor, induced by an external change of the driver parameter $\alpha$. 
We consider changes from $\alpha=3.16$ to $\alpha=3.79$ at $t=250$ time units. We have assured that at time $t=0$ the dynamics is on the attractor, by having employed a preceding integration for the long integration time of $10^6$ time units. We choose $m=100$ snapshots to reconstruct the dynamics separated by $\delta t= \Delta t=0.01$. Results are shown in Figure~\ref{f.KS_alpha}. 
It is seen that the number of POD modes required for accurate reconstruction is higher for the larger value of $\alpha$, consistent with the analytical results that the attractor dimension increases with $\alpha$ \cite{Temam}. In this example the change in the attractor is not sufficiently strong to prompt a prolonged relaxation towards the new attractor, which occurs on a fast time scale that cannot be resolved by the given sampling time $\delta t$. Instead the dynamics is well recovered both before and after the transition. The transition is detected here in the sudden change of POD modes required to perform reliable reconstruction. Figure~\ref{f.KS_alpha_meanvar} shows the reconstruction error at fixed value of $r$. We chose $r=3$ for which the reconstruction error has significantly dropped for $\alpha=3.16$ and an accurate reconstruction is achieved. It is seen that the transition is reflected in the change of the statistics of the reconstruction error at fixed value of $r$. In particular, the mean and the variance of the reconstruction error $\E$, estimated separately for times $t\le 250$ and for times $t>250$, corresponding to $\alpha=3.16$ and $\alpha=3.79$, respectively, are plotted as a function of $r$ and exhibit different orders of magnitude as well as different decay properties upon increasing $r$. For values $r>10$ the error becomes very large due to eigenvalues being wrongly estimated to lie outside the unit circle.

%


\subsection{Detecting the inter-decadal changes in the North Atlantic Oscillation}
\label{sec.NCEP_NAO}
\begin{figure}[htbp]
	\centering
	\includegraphics[width = 0.7\columnwidth]{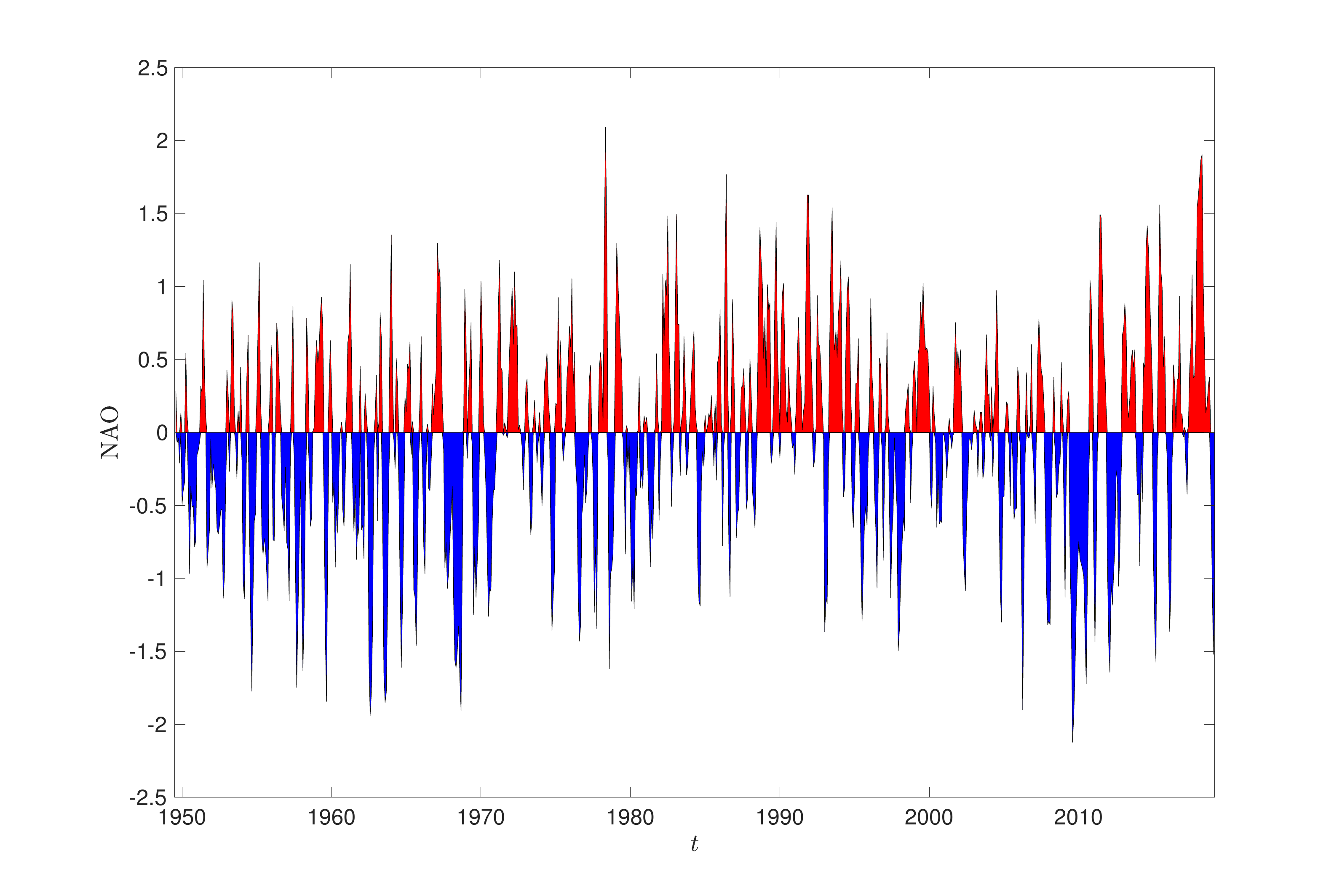}
	\caption{Standardized $3$-months running average NAO Index from January 1950 to July 2019, using the historical data set provided in \cite{CPC}.}
	\label{f.NAOIndex}
\end{figure}

The North Atlantic Oscillation is a major source of low frequency variability in the Northern Hemisphere (defined as variability on time scales larger than 10 days) \cite{HurrellEtAl13,Dijkstra}. The NAO involves changes of the locations of the storm tracks, separating air masses between the Arctic and the subtropical Atlantic, with major impact on heat and moisture transport. These changes are reflected in the NAO index which quantifies the difference of atmospheric pressure at sea level between the Icelandic Low and the Azores High. NAOs are coarsely distinguished in a positive and a negative NAO index phase, where the former is associated with more zonal flow patterns over Europe and the latter is associated with a higher frequency of atmospheric blocking events over the North Atlantic. In Figure~\ref{f.NAOIndex} we show the NAO index (as a $3$ months running average) from January 1950 until July 2019 using the publicly available historical data from the National Oceanic and Atmospheric Administration (NOAA) Climate Prediction Centre (CPC) \cite{CPC}. The NAO has experienced significant changes in the past decades. In particular, the NAO has changed from a predominantly negative phase during the period 1950-1970 to a predominantly positive phase \cite{HurrellVanLoon97}. This regime transition has been linked to a change in forecast skill, with improved forecast skill in the positive phase after 1970 \cite{WeisheimerEtAl17}. Since NAO transitions are believed to be a measure of zonality of the atmospheric flow, meteorologists studied changes in the frequency of atmospheric blocking events, in which the zonal flow splits around a high pressure field which remains nearly stationary on a time scale of several days up to weeks. To identify blocking events and to systematically identify metastable atmospheric regimes in high-dimensional data sets physics-informed indicators were employed \cite{LejenaesOkland83,TibaldiMolteni90}, as well as sophisticated data-driven approaches using clustering algorithms such as k-means \cite{MoGhil88,KondrashovEtAl14} and non-stationary time series analysis methods \cite{Franzke09,FranzkeEtAl09,OKane13a}. More recently methods from dynamical systems theory have been employed. In particular, covariant Lyapunov vectors and unstable periodic orbits were shown to capture blocking events \cite{SchubertLucarini16,LucariniGritsun19}. More closely related to our approach of DMD, methods based on the transfer operator, the formal $L_2$-adjoint of the Koopman operator, and the behaviour of its point spectrum were used to study transitions to individual blocking event \cite{TantetEtAl15}.

\begin{figure}[htbp]
	\centering
	\includegraphics[width = 0.5\columnwidth]{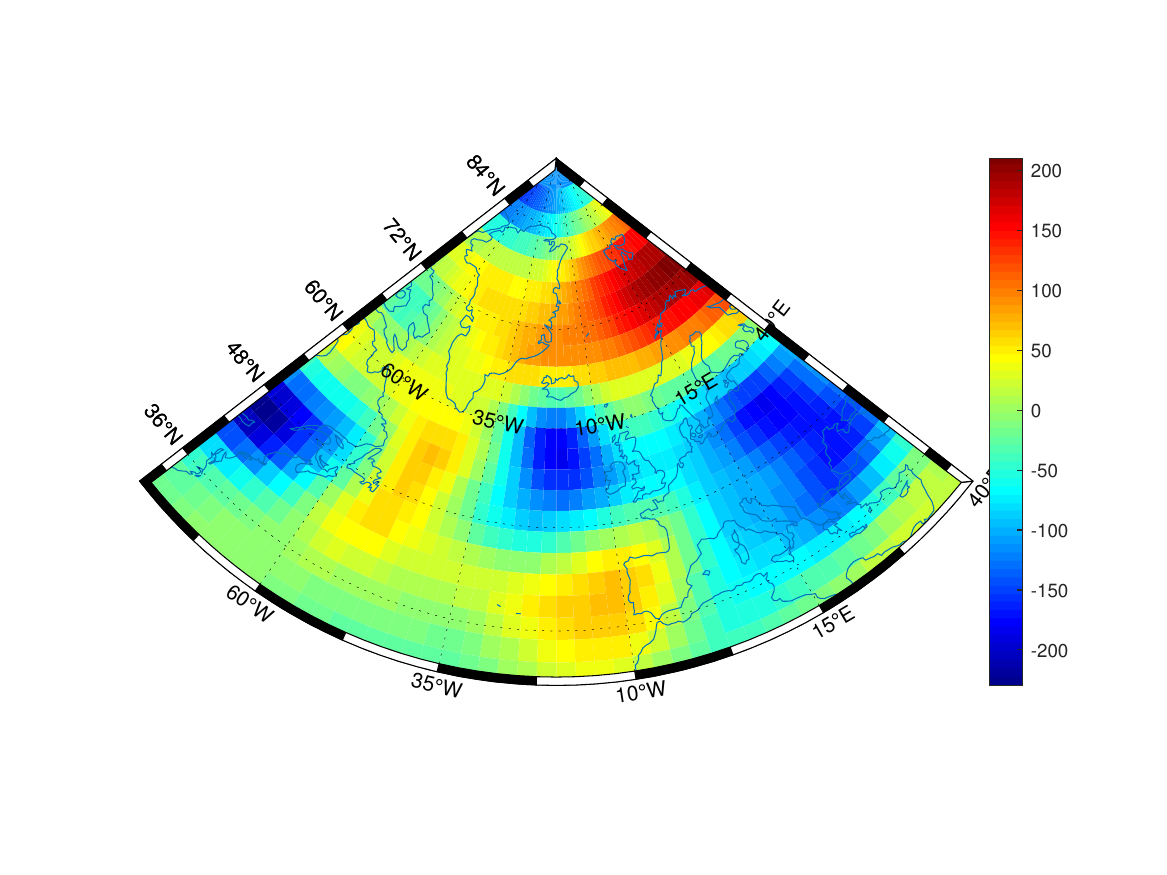}\\
	\vspace{0.1cm}
	\includegraphics[width = 0.5\columnwidth]{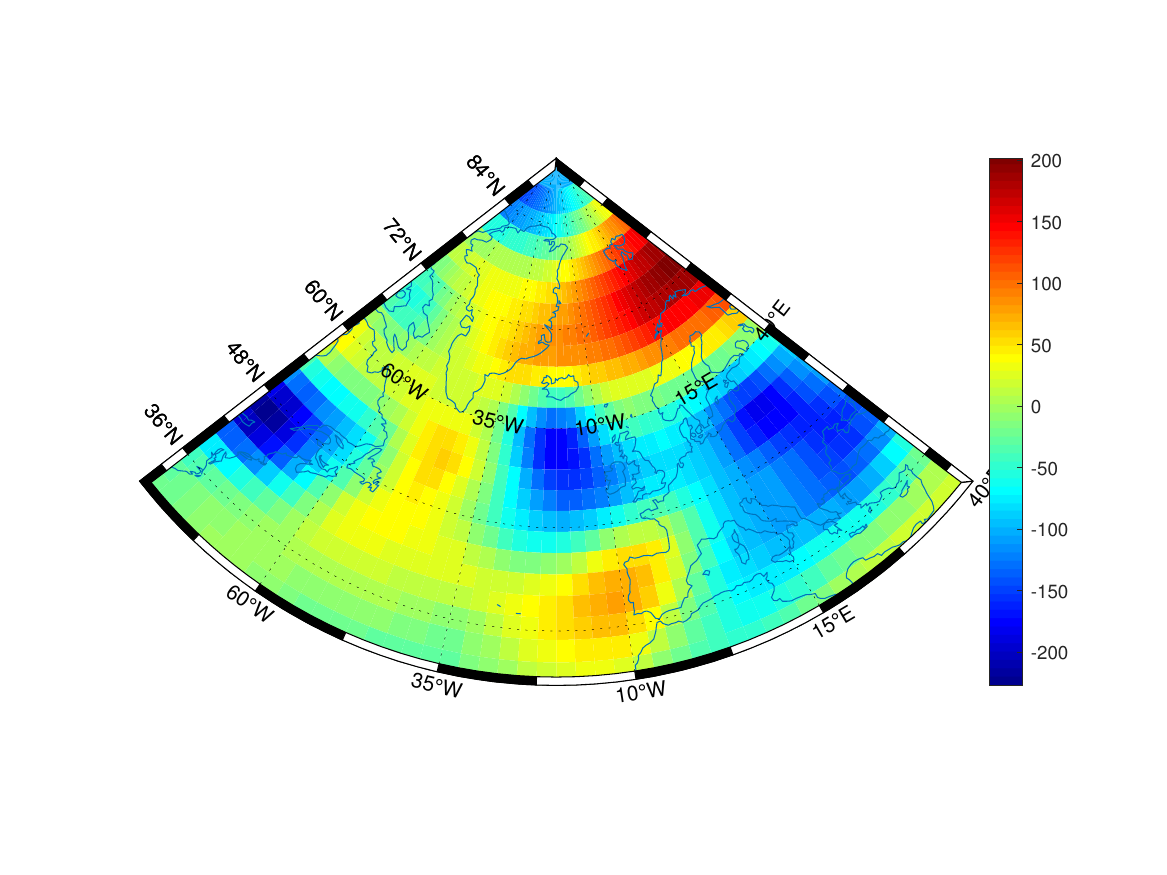}
	\caption{Anomaly of the $500$hPa geopotential height on July 5 1960 at noon in the Northern Hemisphere. Top: Reanalysis data. Bottom: DMD reconstruction for $r=6$ for the same time initiated at July 1 1960 at noon. Figures were made using the software by \cite{mproj}.}
	\label{f.NHdata}
\end{figure}

We shall use the DMD reconstruction error diagnostics to identify NAO transitions in NCEP/NCAR (National Centers for Environmental Prediction/National Center for Atmospheric Research) reanalysis data \cite{NCEPdata}. We use their 6 hourly reanalysis data of the $500$hPa geopotential height covering the temporal period 1948--2017 for the spatial region of the Northern Hemisphere in the region 
[$30$N--$90$N, $80$W--$40$E] with a spatial resolution of $2.5^\circ\times 2.5^\circ$. We consider only anomalies with respect to the climatological mean without detrending. Figure~\ref{f.NHdata} shows a snapshot of the geopotential height anomalies at July 5 1960 at noon.

\begin{figure}[htbp]
	\centering
	\includegraphics[width = 0.5\columnwidth]{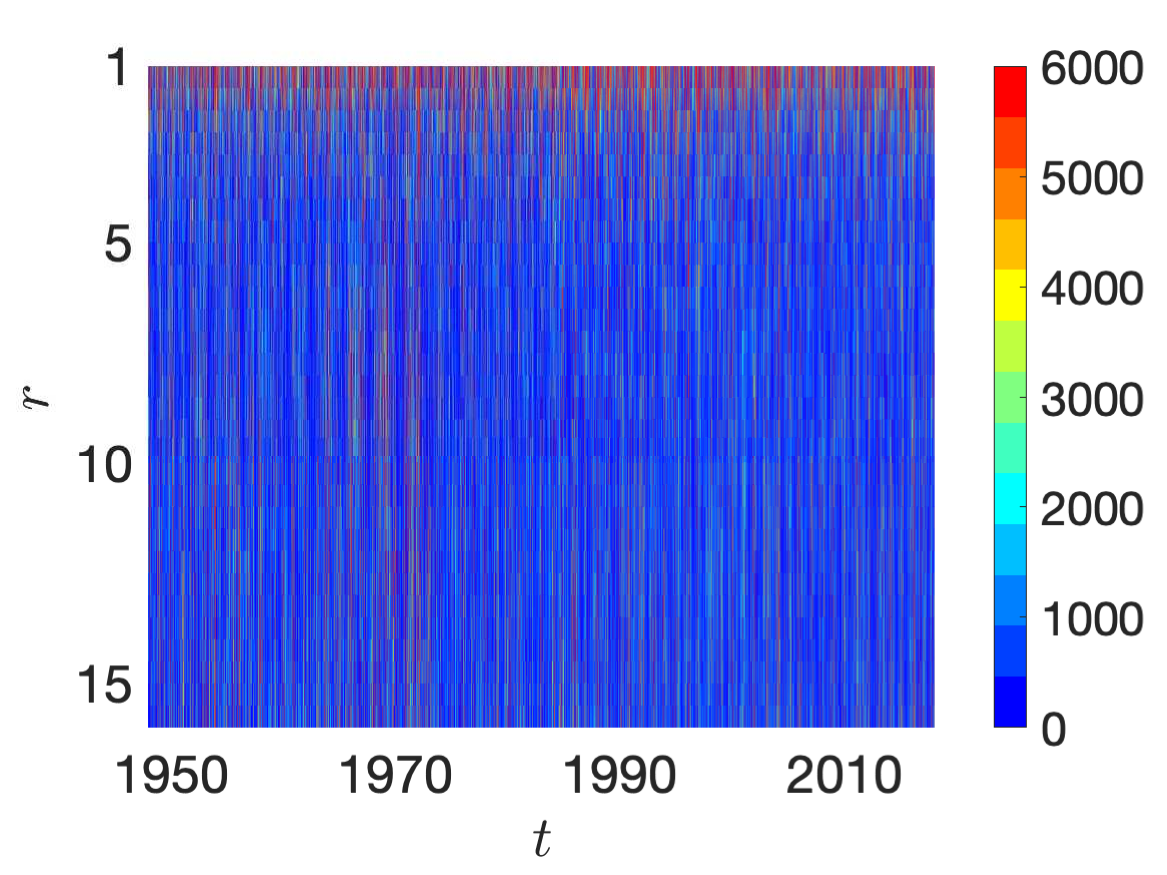}\\
	\vspace{0.1cm}
	\includegraphics[width = 0.5\columnwidth]{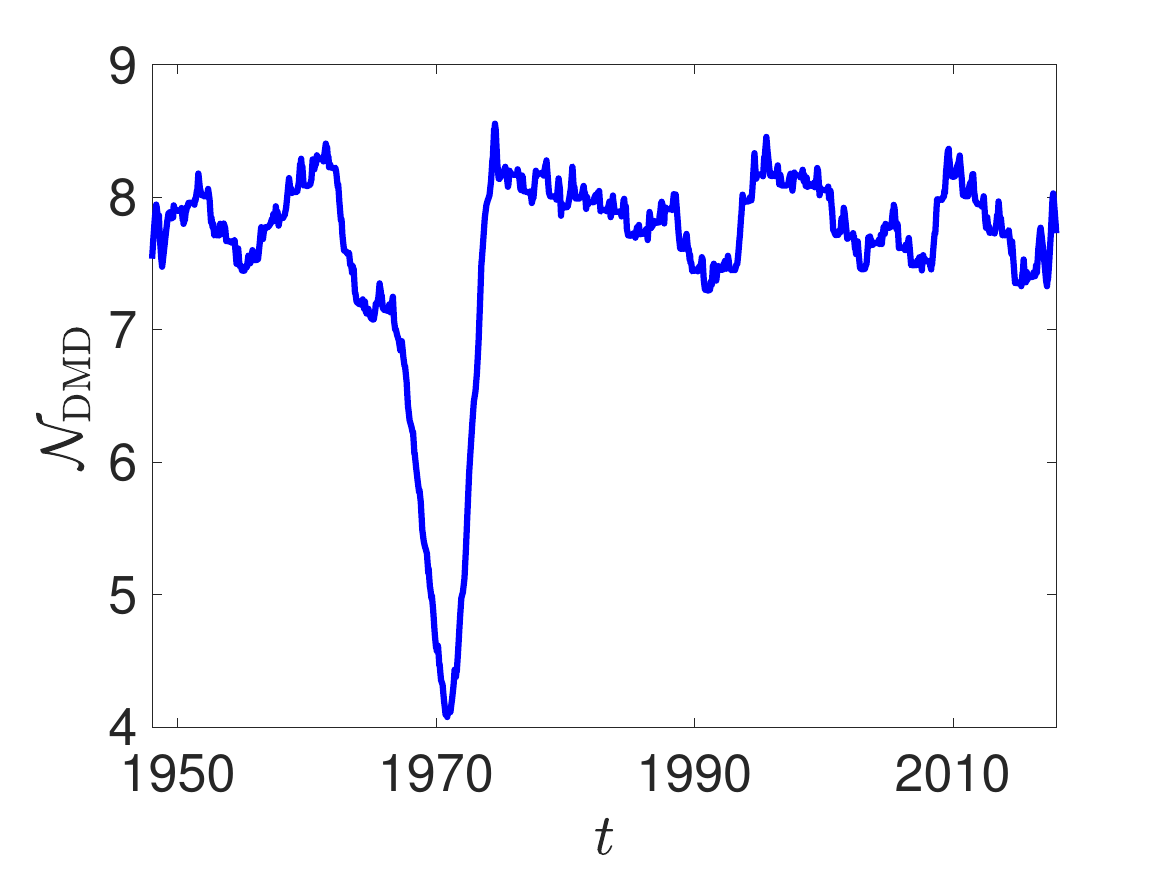}
	\caption{Diagnostics of the reconstruction error for the reanalysis data of the $500$hPa geopotential height anomalies in the Northern Hemisphere. Top: Reconstruction error $\E(t,r)$. Bottom: Average number of instances of good reconstruction with reconstruction errors ${\mathcal{E}}_{\rm{DMD}}\le 400$ for $7\le r \le 16$.}
	\label{f.NHE}
\end{figure}

The time intervals for the Koopman analysis are $\Delta t=\delta t=6$ hours and we choose the geopotential height at $500$hPa at every grid point of the whole domain as our $n=1225$-dimensional observable for the DMD analysis. The DMD reconstruction is computed for a synoptic time window of $4$ days, which corresponds to $m=16$. We have also performed an analysis using $m=40$, which corresponds to a characteristic time scale of atmospheric blocking events, with qualitatively similar results. In Figure~\ref{f.NHdata} an instance of a good reconstruction error with $\E$ is shown with $\E\approx 88$. The DMD reconstruction relies on the singular value decomposition. It is well known that for observations which are noise contaminated to some degree, only the first singular values and their associated left and right eigenvectors are reliable, and one should truncate the singular value decomposition; recall the very large reconstruction errors for $r\approx m$ in the case of the Kuramoto-Sivashinsky equation discussed in Section~\ref{sec.KS}. For additive measurement noise, Gavish and Donoho \cite{GavishDonoho14} provide a criterion for an optimal truncation. The criterion yields for our set-up an optimal truncation at $r=6$. Figure~\ref{f.NHE} shows the reconstruction error $\E$ for the whole period 1949--2017 for $r\le 16$. We observe an increased error for values of $r$ around the optimal truncation value $r=6$ for the years around 1970. To quantify this we plot $\mathcal{N}_{\rm{DMD}}$, the moving average of the number of $r$-values with acceptable reconstruction accuracy with $\E \le 400$ between $7\le r \le 16$ (the average is performed over three years). Note that large values of $\mathcal{N}_{\rm{DMD}}$ imply good reconstruction and that by construction $\mathcal{N}_{\rm{DMD}}\le 16-7+1=10$. There is a clear dip in $\mathcal{N}_{\rm{DMD}}$ around 1970, suggesting transitory dynamics, consistent with the observed transition from a predominantly negative NAO phase to a predominantly positive NAO phase \cite{HurrellVanLoon97}. 

Besides this large-scale regime change in the 1970s, our diagnostics $\N$ may also be able to detect transitions occurring on smaller temporal scales. The NAO index exhibits a pronounced period of negative NAO around the year 2010, as shown in Figure~\ref{f.NAOIndex}. Our transition diagnostics $\N$ exhibits a peak at the same time (albeit of much less magnitude than the dip around the 1970s). Similarly, the NAO index is strongly positive around 1990, which correlates with a dip in the value of $\N$ at that time. This is suggestive that $\N$ is able to distinguish zonal from blocked states, with peaks in $\N$, indicating improved DMD reconstruction, corresponding to predominantly zonal flow and dips in $\N$, indicating poorer DMD reconstruction, to  predominantly blocked states. This is consistent with the fact that DMD analysis works well for linear dynamics since the algorithm corresponds to linear least square fits of the dynamics (cf (\ref{e.XXpseudo})), and that naturally linear dynamics is associated with better weather forecasting skill.

We caution that our method does not hint to any causal physical mechanisms responsible for a detected regime change. It is by no means clear if the regime changes, which were detected with our test, are in any way related to changes in the NAO or to other physical mechanisms, or even due to changes in the observational system used for the reanalysis. The diagnostics is only able to quantify the ability of DMD to produce a reliable reconstruction of the observations; failure to do so and changes in the performance of DMD may have numerous reasons, and for transient dynamics we argue that it is caused by the dynamics evolving on time scales faster than the sampling time or by a varying degree of linearity of the underlying dynamics.


 
\subsection{Detecting inter-decadal changes in the large-scale circulation of the Southern Hemisphere}
\label{sec.NCEP}
As the Northern Hemisphere atmospheric dynamics the Southern Hemisphere atmospheric dynamics and its climate has experienced significant changes in the past decades. In particular, the frequency of blocking events has decreased significantly around the mid 1970s and has given way to a more zonal flow regime \cite{OKane13a,OKane13b}. This regime change was shown to be likely caused by anthropogenic ${\rm{CO}}_2$ emissions \cite{FranzkeEtAl15,FreitasEtAl15}. 

\begin{figure}[htbp]
	\centering
	\includegraphics[width = 0.5\columnwidth]{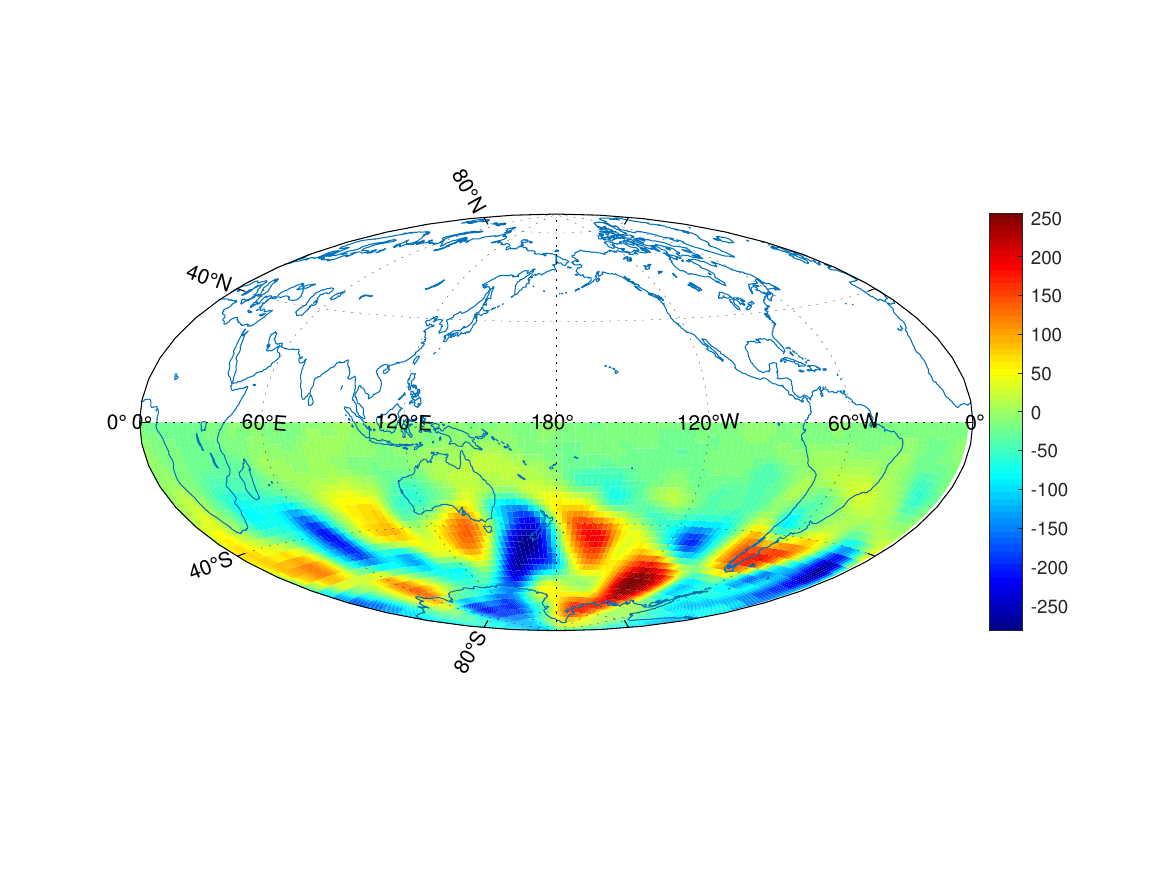}\\
	\vspace{0.1cm}
	\includegraphics[width = 0.5\columnwidth]{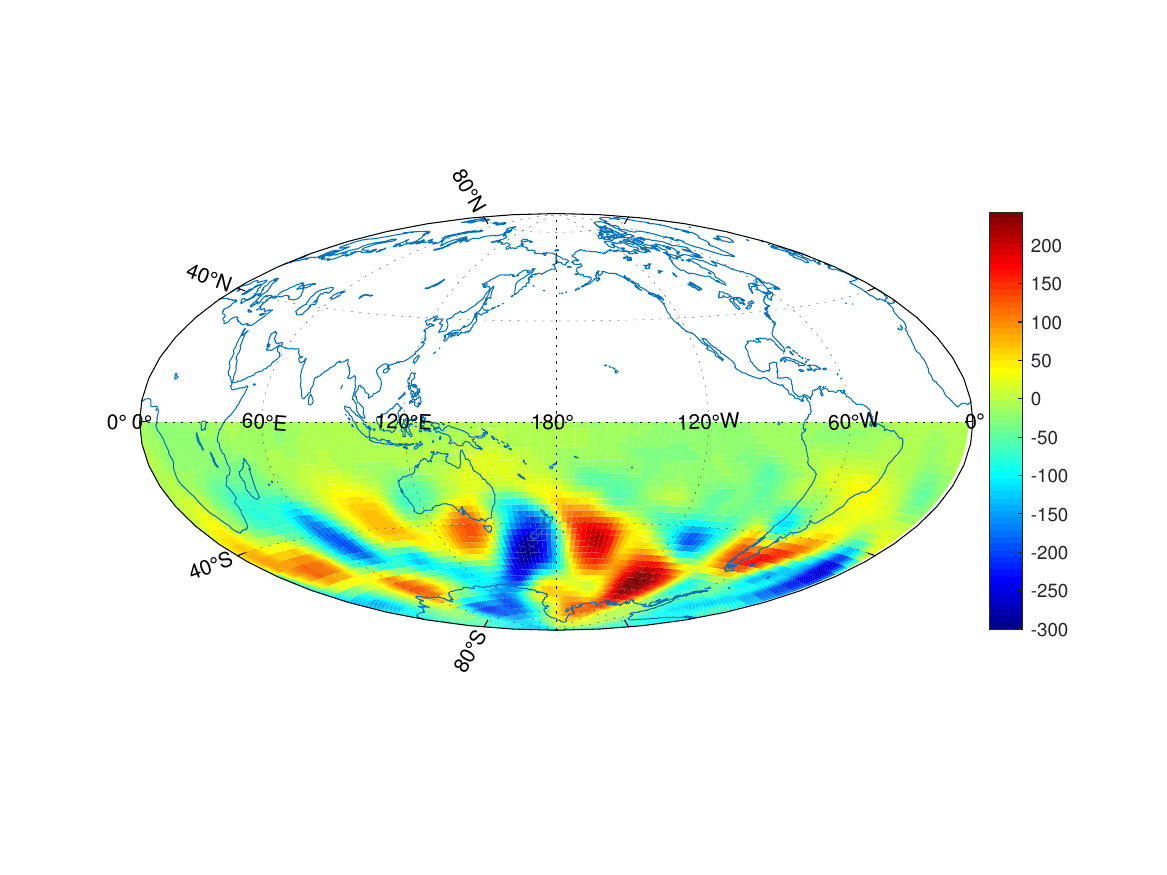}
	\caption{Anomaly of the $500$hPa geopotential height on July 5 1960 at noon in the Southern Hemisphere. Top: Reanalysis data. Bottom: DMD reconstruction for $r=6$ for the same time initiated at July 1 1960 at noon. Figures were made using the software by \cite{mproj}.}
	\label{f.SHdata}
\end{figure}

We employ again the reconstruction error diagnostics to detect this change in the NCEP/NCAR reanalysis data \cite{NCEPdata}. We use again the 6 hourly reanalysis data of the $500$hPa geopotential height covering the temporal period 1948--2017 but consider now the spatial region of the full Southern Hemisphere. As before, we consider anomalies with respect to the climatological mean without detrending. We provide in Figure~\ref{f.SHdata} a snapshot of the geopotential height anomalies on  
July 5 1960 at noon.

The time intervals for the Koopman analysis are $\Delta t=\delta t=6$ hours, and we choose the geopotential height at $500$hPa at every grid point of the whole domain as our $n=5328$-dimensional observable for the DMD analysis. We choose a reconstruction window of $m=16$, corresponding to a characteristic time synoptic scale of $4$ days. We found qualitatively similar results when extending the time window to $10$ days, resolving the typical time scale for atmospheric blocking events. The optimal truncation criterion of Gavish and Donoho \cite{GavishDonoho14} yields $r=6$ for this set-up.

\begin{figure}[htbp]
	\centering
	\includegraphics[width = 0.5\columnwidth]{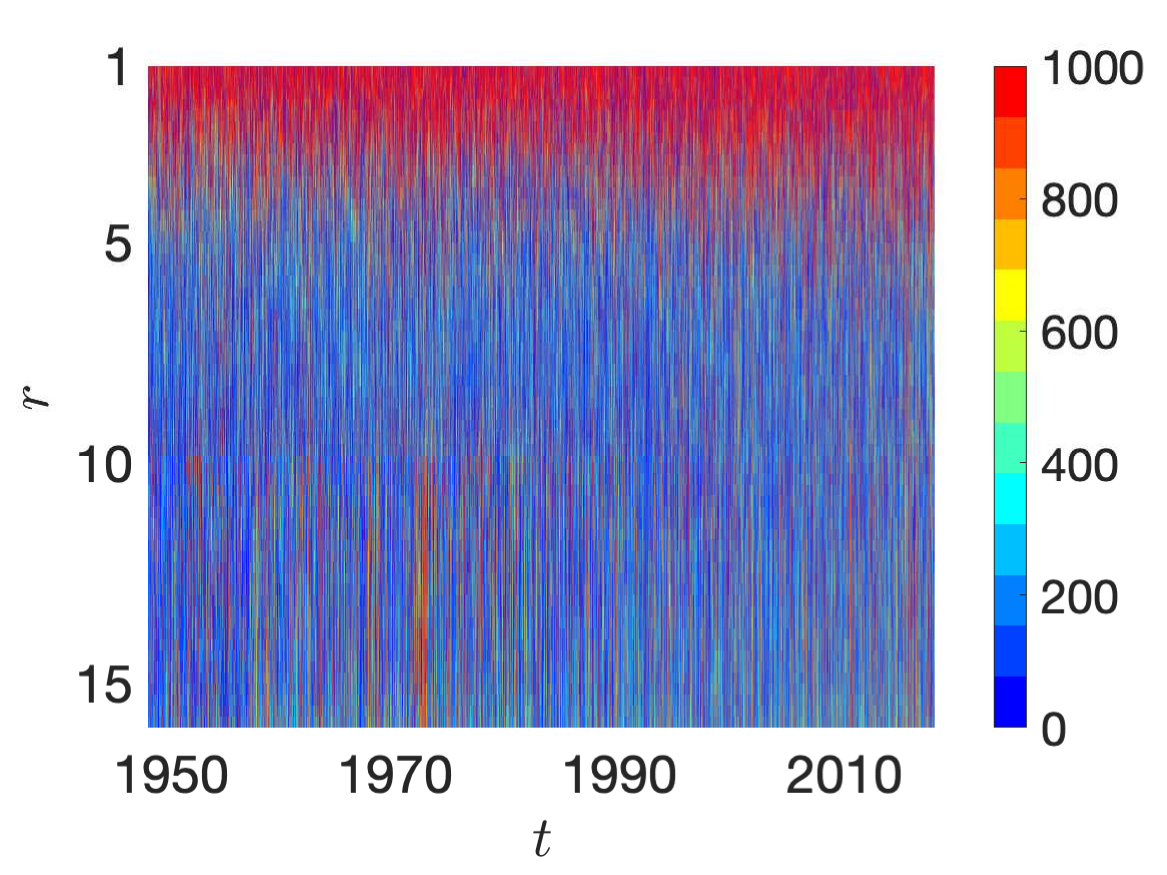}\\
	\vspace{0.1cm}
	\includegraphics[width = 0.5\columnwidth]{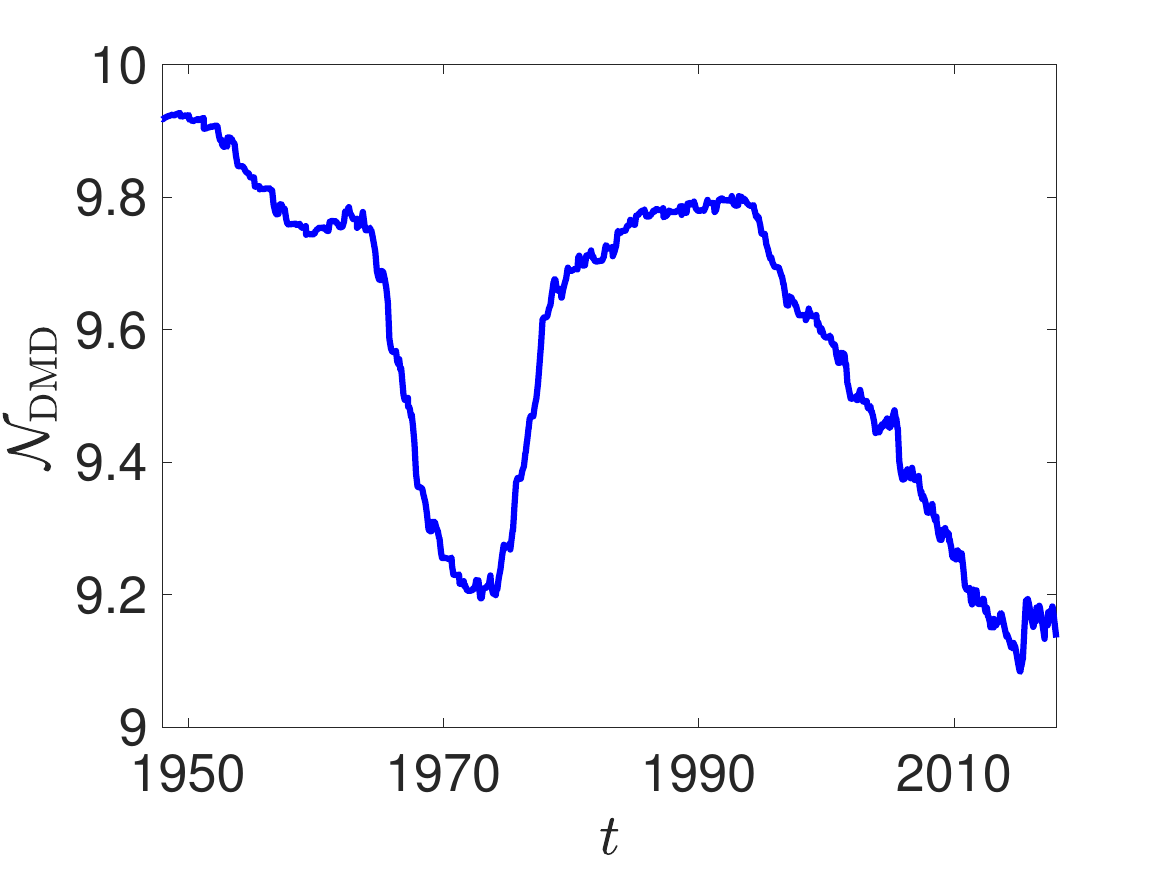}
	\caption{Diagnostics of the reconstruction error for the reanalysis data of the $500$hPa geopotential height anomalies in the Southern Hemisphere. Top: Reconstruction error $\E(t,r)$. Bottom: Average number of instances of good reconstruction with reconstruction errors ${\mathcal{E}}_{\rm{DMD}}\le 400$ for $7\le r \le 16$.}
	\label{f.SHE}
\end{figure}

Figure~\ref{f.SHdata} shows an instance of a good reconstruction error $\E\approx 322$ for $r=6$. The reconstruction error $\E$ for the whole period 1948--2017 is shown for $r\le 16$ in Figure~\ref{f.SHE}. We observe an increased error around the optimal reconstruction error $r=6$ for the years around 1970. To quantify this we evaluate again the moving average  $\mathcal{N}_{\rm{DMD}}$ of the number of $r$-values with acceptable reconstruction accuracy with $\E \le 400$ between $7\le r \le 16$ (the average is performed over ten years). By construction $\mathcal{N}_{\rm{DMD}}\le 16-7+1=10$. There is a clear dip in $\mathcal{N}_{\rm{DMD}}$ around 1970, suggesting transitory dynamics, consistent with the observed regime change associated with a decrease of blocking instances \cite{OKane13a,OKane13b}. Noticeable, the reconstruction error increases continuously after 1990. This may be due to further hitherto unidentified slow transitory dynamics or to a higher degree of nonlinearity of the dynamics.

\section{Discussion}
\label{sec.discussion}
We proposed a data-driven method to detect eventual regime changes and transient dynamics in time series. Our method is based on the Koopman operator and on approximating its eigenmodes using dynamic mode decomposition. Our approach exploits the fact that transient non-equilibrium dynamics typically evolves on fast time scales and that for a given temporal sampling interval DMD may fail to reconstruct the dynamics during periods of fast relaxation towards an attractor. Moreover, transient non-equilibrium dynamics may explore the full-dimensional state space, as opposed to equilibrium dynamics which is typically supported on a lower dimensional attractor. We use the reconstruction error between the observations and their DMD approximations as a diagnostic tool to analyze the dynamics and probe transient behaviour and regime changes. We studied the behaviour of the reconstruction error for different values of the low-rank approximation parameter $r$, and showed that one can use the reconstruction error at fixed values of $r$ as well as the dependency of its statistics, such as mean and variance, to infer transitions. We have illustrated the potential of our method to detect transitions and regime changes in artificially prepared data from a chaotic partial differential equation as well as in reanalysis data of the Southern and Northern Hemisphere atmospheric climate data. Our method is cost-effective and is able to detect regime changes and transitory dynamics in both cases. The diagnostics we propose crucially depends on the transition evolving on a time scale faster than the resolution time scale $\delta t$ used to estimate the Koopman operator. If the transition time scale is of the same order as the equilibrium dynamics then the transition dynamics is not reflected in the magnitude of the reconstruction error, even using subsampling of the time series to increase the time step $\delta t$. We have shown, however, that changes in the statistics of the reconstruction error may still be used to detect transitions and a qualitative change of the equilibrium dynamics, even when the reconstruction errors are small for the whole time of the observations.\\ 

It is important to stress that our method cannot detect regime transitions per se but only those which occur on time scales which are faster than the sampling time. Moreover, our method is entirely data-driven and does not allow for the attribution of causal physical mechanisms responsible for a detected regime change. Given these caveats we cannot recommend to employ our method in isolation, but rather to concurrently consult physics-based analyses to draw informed conclusions.\\

We were concerned here with identifying periods of transient dynamics and regime changes. In future work it may be interesting to see whether DMD is able to identify precursors of transitions, and thereby to study if DMD is able to predict transitions. The choice of observables will then be of importance. Furthermore, we considered here dissipative dynamical systems; recently finite-rank approximations of Koopman operators were discussed for measure preserving dynamical systems \cite{GovindarajanEtAl19} and it would be interesting to see if one can devise similar diagnostics as proposed here in this context. 
\section*{Acknowledgments}
We thank Terry O'Kane for alerting us to the NCEP Reanalysis data and the change in circulation patterns in the Southern Hemisphere, and Joshua Dorrington for alerting us to the change in the NAO. We thank P\'eter Koltai for comments on an earlier draft of the paper.







\end{document}